\documentclass[a4paper,fleqn,usenatbib]{mnras}
\pdfoutput=1
\usepackage[T1]{fontenc}
\usepackage{ae,aecompl}

\usepackage{graphicx}
\usepackage{amsmath}
\usepackage{amssymb}
\usepackage[ddmmyyyy]{datetime}
\usepackage{grffile}    
\usepackage{blindtext}  
\usepackage{ulem}

\graphicspath{{figures/}}
\newcommand{\dint}{\mathrm{d}}
\newcommand{\rew}{\mathrm{REW}}
\newcommand{\muv}{M_\mathrm{UV}}

\hypersetup{breaklinks}

\newcommand{\colb}{blue}
\newcommand{\colg}{green}
\newcommand{\colr}{red}

\title[High-z LAE Modelling]{Modelling the observed luminosity 
function and clustering evolution of Lyman-$\alpha$ emitters: 
growing evidence for late reionization}

\author[L. Weinberger et al.]{
Lewis H. Weinberger$^{1}$\thanks{Email: lewis.weinberger@ast.cam.ac.uk},
Martin G. Haehnelt$^{1}$ and
Girish Kulkarni$^{2}$
\\
$^{1}$ Institute of Astronomy and Kavli Institute for Cosmology, University
of Cambridge, Madingley Road, Cambridge CB3 0HA, UK \\
$^{2}$ Department of Theoretical Physics, Tata Institute of Fundamental Research, 
Homi Bhabha Road, Mumbai 400005, India
}

\date{Accepted XXX\@. Received YYY;\ in original form ZZZ}
\pubyear{2018}

\begin{document}%
\label{firstpage}
\pagerange{\pageref{firstpage}--\pageref{lastpage}}
\maketitle

\begin{abstract}
We model the high redshift ($z>5$) Lyman-$\alpha$ emitting (LAE) galaxy population 
using the empirical rest-frame equivalent width distribution. We calibrate to 
the observed luminosity function and angular correlation function at $z=5.7$ as 
measured by the SILVERRUSH survey.
This allows us to populate the high-dynamic-range 
Sherwood simulation suite with LAEs, and to calculate the transmission of their Ly $\alpha$ 
emission through the inter-galactic medium (IGM). 
We use this simulated population to explore the
effect of the IGM on high-redshift observations of LAEs, and make predictions
for the narrowband filter redshifts at $z=6.6$, $7.0$ and $7.3$.  Comparing our 
model with existing observations, we find a late reionization is suggested, 
consistent with the recent low optical depth derived from the cosmic microwave 
background (CMB) by the Planck 
collaboration and the opacity fluctuations in the Ly $\alpha$ forest. 
We also explore the role of the circum-galactic medium (CGM) and the larger
volume of gas which is infalling onto the host halo
versus the IGM in attenuating the Ly $\alpha$ signal, finding  that a significant 
fraction of the attenuation is due to the CGM and infalling gas, 
which increases towards the end 
of reionization, albeit with a large scatter across the mock LAE population.
\end{abstract}

\begin{keywords}
galaxies: high-redshift - galaxies: evolution -
reionization - intergalactic medium - cosmology: theory
\end{keywords}



\section{Introduction}

Observational studies of Lyman-$\alpha$ emitting galaxies (LAEs) at high redshifts,
$z \geq 5$, have now amassed a considerable population of objects that
can be used to learn much about the reionization era and galaxy evolution. 
These include widefield narrowband surveys such as \citet{2018arXiv180505944I,
2017arXiv170501222K,2017arXiv170302985Z,2017ApJ...837...11B,2017arXiv170302501O,
2016MNRAS.463.1678S,
2015MNRAS.451..400M,2014ApJ...797...16K,2011ApJ...734..119K,2010ApJ...723..869O,
2008ApJS..176..301O}, probing redshifts $z=$ 5.7, 
6.6, 7.0 and 7.3 using filters on instruments such as the SuprimeCam and 
HyperSuprimeCam of the Subaru telescope \citep{2002PASJ...54..833M,
2012SPIE.8446E..0ZM}. 
Spectroscopic studies such as 
\citet{2018MNRAS.tmp.1569M,2017MNRAS.472..772M,2017MNRAS.471.3186D,
2014ApJ...788...74S,2014ApJ...793..113P,2012ApJ...744...83O,2012ApJ...744..179S,
2011ApJ...728L...2S,2010MNRAS.408.1628S,2006ApJ...648....7K}
have allowed detailed confirmation of some of the most interesting objects,
including CR7 \citep{2015ApJ...808..139S,2017arXiv171008422S} and COLA1 
\citep{2016ApJ...825L...7H,2018arXiv180511621M}, 
as well as a better understanding of how LAEs fit into a bigger picture of 
high-$z$ galaxies.

\begin{figure*}
    \includegraphics[width=2\columnwidth]{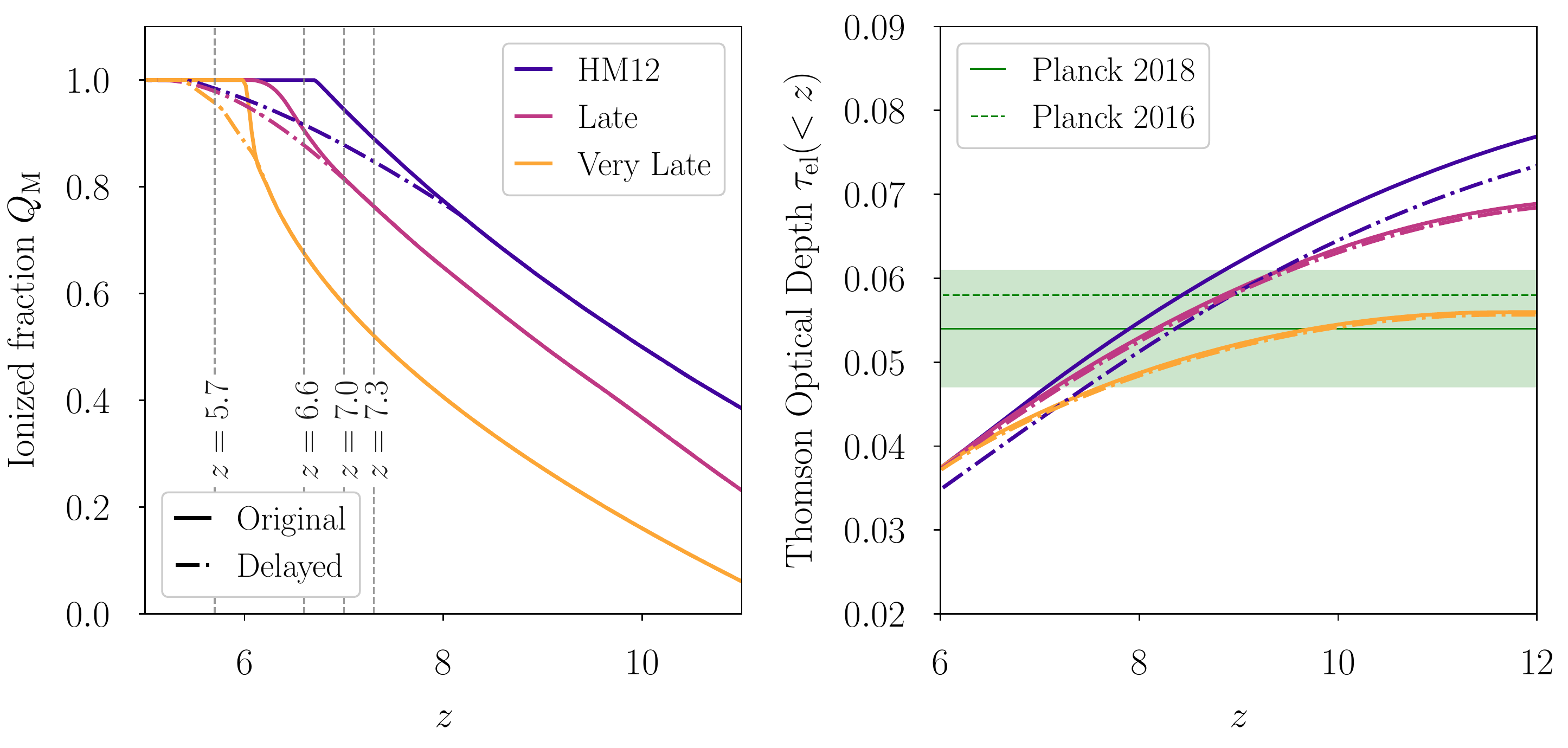}
    \caption{Left: Mass-averaged global ionized fraction for the reionization 
    histories tested in this work, labelled HM12 (blue)
    Late (purple), Very Late (orange). The solid lines show the original
    models, whilst the dash-dotted lines are modified to have a delayed end
    to reionization.
    The HM12 model reionizes completely by
    $z=6.7$, whilst the Late/Very Late models finish by $z=6$, and the Delayed models
    finish at $z=5.3$. These six histories
    allow us to bracket the range of possible reionization paths that are
    constrained by CMB and Ly $\alpha$ forest measurements. Right: the
    Thomson optical depth to electron scattering as a function of redshift for
    the different reionization histories. For comparison the  
 	\citet{2018arXiv180706209P} and \citet{2016A&A...596A.108P} 
    CMB results are shown in green (with shading indicating the 1 $\sigma$
    uncertainty).}
    \label{fig:calib}
\end{figure*}

LAEs have been established as a key tool for understanding the progress of 
reionization at $z\geq6$. Given the resonant scattering of Ly $\alpha$ by 
neutral hydrogen in the intergalactic medium (IGM) and circumgalactic medium 
(CGM) \citep{1965ApJ...142.1633G,2000ApJ...542L..69M}, the visibility of LAEs should
decrease as observations probe further into the reionization era \citep{2007MNRAS.377.1175D}.
This attenuation can be seen in the redshift evolution of both the luminosity
function \citep[e.g.][]{2017arXiv170501222K} and the clustering signal 
\citep[e.g.][]{2017arXiv170407455O}.
Studies at lower redshifts are also important for understanding the ionizing photon
budget, for example \citet{2018MNRAS.477.2098N} used LAEs to understand ionizing
photon escape fractions, to determine if LAEs could have played a 
significant role in (re)ionizing their surrounding IGM. Green Pea galaxies,
low-redshift analogs of LAEs, present a further avenue for understanding
the properties of LAEs \citep{2016ApJ...820..130Y}.

There has been extensive theoretical modelling of LAEs, considering both the
`intrinsic' Ly $\alpha$ emission properties of galaxies as a result of
radiative transfer within the halo
\citep[such as][]{2016ApJ...826...14G,2014PASA...31...40D,2010ApJ...716..574Z,2010MNRAS.401.2343D,
2009ApJ...704.1640L,1967ApJ...147..868P} as well as the
effect of further attenuation by resonant scattering with neutral gas in the CGM and IGM 
\citep[such as][]{2018arXiv180607392L,2018arXiv180101891M,2018arXiv180100067I,
2018MNRAS.tmp.1485W,2017ApJ...839...44S,2016MNRAS.463.4019K,
2015MNRAS.450.4025H,2015MNRAS.446..566M,2015MNRAS.452..261C,
2014MNRAS.441.2861H,2014MNRAS.444.2114J,2013MNRAS.428.1366J,
2013MNRAS.429.1695B,2011ApJ...728...52L,
2011MNRAS.414.2139D,2009MNRAS.400.2000D}. In particular
these two regimes are often modelled differently: the escape of Ly $\alpha$
photons from within a galaxy is frequently treated using full radiative transfer
post-processing of high resolution hydrodynamic simulations \citep[e.g.][]{2006A&A...460..397V}, whilst the 
scattering in the IGM can be well approximated using $e^{-\tau}$ models
\citep[e.g.][]{2011ApJ...728...52L}. 
On top of this it is important to model how LAEs form part of the
wider galaxy population; \citet{2012MNRAS.419.3181D} for example modelled LAEs as a subset of the
Lyman break galaxy (LBG) population with a $M_{UV}$-dependent distribution of
Ly $\alpha$ equivalent widths. Given a mapping between host halo mass and galaxy
UV luminosity, it is possible to employ such a model to fit the UV and Ly $\alpha$
luminosity function evolution, whilst also being consistent with observed equivalent
width distributions. There is however some degeneracy in the mapping from mass
to $M_{UV}$, which can be broken by considering the spatial clustering of LAEs
\citep{2009ApJ...695..368L}. It is possible to fit all these observational 
constraints if a duty cycle is employed \citep{2010ApJ...714L.202T}.

In \citet{2018MNRAS.tmp.1485W} we employed large, high-dynamic-range hydrodynamical simulations 
of the IGM in combination with a semi-analytic reionization model in order to
calculate the transmission of Ly $\alpha$ from LAE host haloes. The effects of
the CGM, self-shielded neutral gas, and host halo mass dependence on the
transmission were explored. In this work we now extend those transmission models
to include an empirically-constrained model for the population of LAEs. 
Our modelling updates previous work and employs large scale 
simulations of the IGM to test the effects of reionization at observable redshifts,
and to predict the LAE luminosity function, angular correlation function and 
rest-frame equivalent width probability distribution simultaneously.
The underlying simulations have been used to model the 21 cm signal during reionization
\citep{2016MNRAS.463.2583K} as well as opacity fluctuations after reionization 
\citep{2018arXiv180906374K}, such that we will be able to perform like-for-like 
comparisons with these other reionization observables in future work.
This allows us to make predictions for the evolution of the Ly $\alpha$ luminosity 
function and clustering signal, which we compare to available observational
data.

This paper is structured as follows: in section \ref{sec:methods} we 
outline how we model the LAE populations using a suite of cosmological
hydrodynamic simulations, then in section \ref{sec:results} we present a comparison
of the predictions of our models with current observed data. In section
\ref{sec:discussion} we discuss the assumptions of our modelling, before concluding in
section \ref{sec:conclusions}.


\section{Methods}
\label{sec:methods}

\subsection{Numerical simulations of the IGM}
In order to model the IGM gas properties we employ the high-dynamic-range 
Sherwood simulation
suite of cosmological hydrodynamic simulations \citep{2017MNRAS.464..897B}, with the same set up as in \citet{2018MNRAS.tmp.1485W}. 
We refer the reader to that work for full details of the simulation and
reionization modelling, but we summarize the salient points below.

The IGM gas properties are derived from the cosmological hydrodynamical simulations,
which were performed using the \textsc{P-Gadget-3} SPH code, a modified version
of the original \textsc{Gadget} branch \citep{doi:10.1111/j.1365-2966.2005.09655.x,
2001NewA....6...79S}. In this work we utilize a
simulation box of side length $L=320$ cMpc/h and particle number $N=2\times2048^3$,
so as to be large enough to capture the clustering signal at large scales.
Dark matter haloes were identified on-the-fly using a friends-of-friends
algorithm\footnote{We note however that sub-haloes were not identified.}. 
The cosmological parameters were set to match the \citet{2014A&A...571A..16P}
results: $h=0.678$, $\Omega_m = 0.308$, $\Omega_\Lambda = 0.692$, 
$\Omega_b = 0.0482$, $\sigma_8 = 0.829$, $n = 0.961$, and $Y_{\mathrm{He}} = 0.24$.
The gas properties were smoothed onto a uniform grid with cell size 
$L_\mathrm{cell}=156.25$ ckpc/h, using the SPH kernel.

\begin{figure}
    \includegraphics[width=\columnwidth]{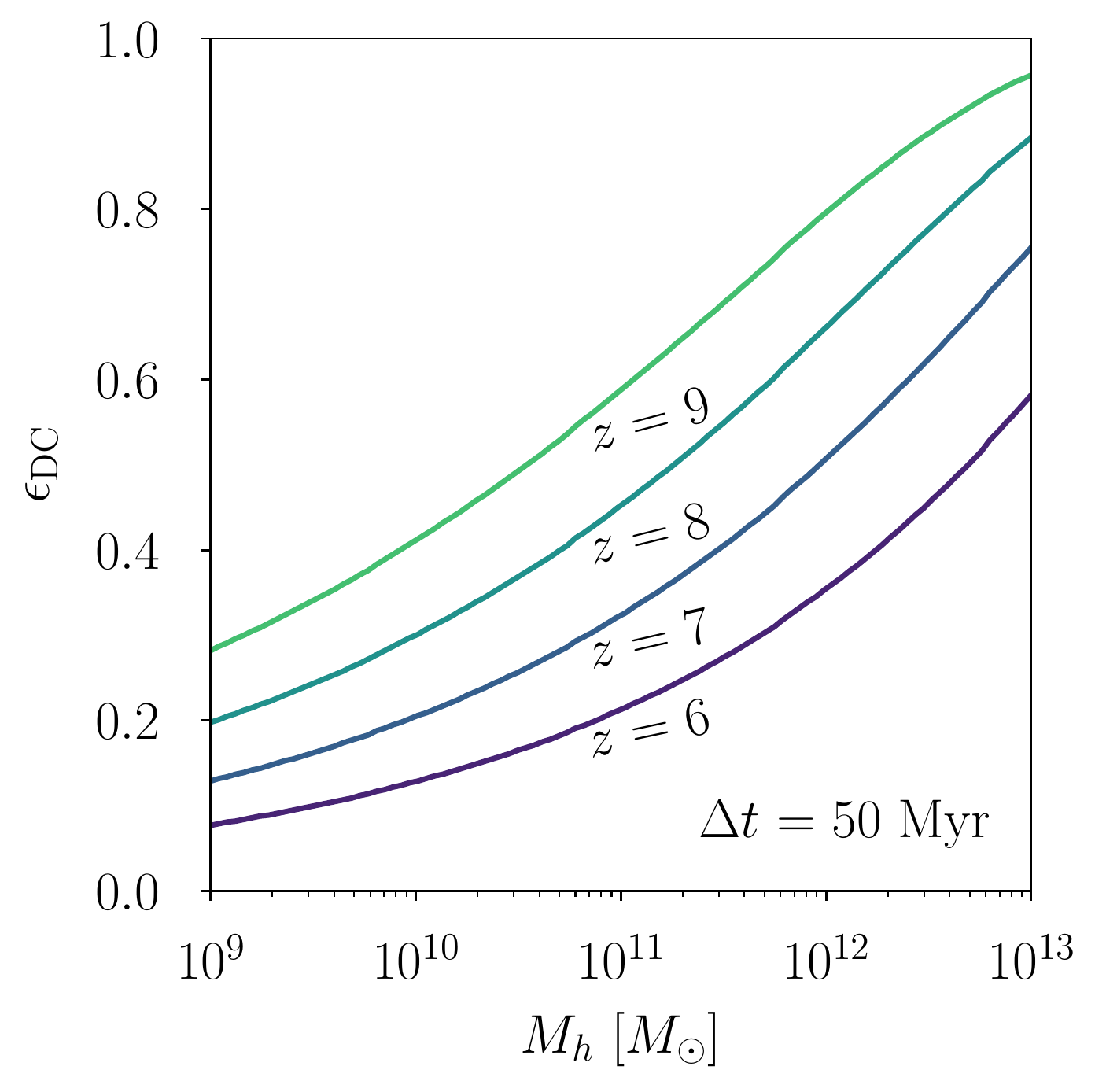}
    \caption{Effective duty cycle defined by \citet{2010ApJ...714L.202T},
    as a function of halo mass and redshift, where
    $\Delta t$ is a free parameter. This weights the abundance matching towards
    haloes that could have formed within $\Delta t$; for example shown here is
    our fiducial choice of $\Delta t = 50$ Myr.}
    \label{fig:edc}
\end{figure}

In order to derive the ionization properties of the gas, a semi-analytic framework
is employed in which the large-scale ionization is modelled using an
excursion set approach \citep{2004ApJ...613....1F,Mesinger:2007pd,2009MNRAS.394..960C,
2011MNRAS.411..955M, 2010MNRAS.406.2421S, 2016MNRAS.457.1550H}
whilst the small-scale ionization is calibrated self-consistently within
the simulation volume \citep{2015MNRAS.452..261C,
2016MNRAS.463.2583K}. In particular we use the self-shielding prescription
of \citet{2017arXiv170706993C}, itself a modification of the 
\citet{2013MNRAS.430.2427R} parametrization, 
in order to accurately model small-scale neutral features. This framework allows
us to test arbitrary reionization histories without needing to explicitly model
the ionizing emissivity of sources. In appendix \ref{appendix:compareRT} we 
compare this excursion set scheme to a full radiative transfer calculation,
and find that our results are largely insensitive to the use of this 
approximation.

As in \citet{2018MNRAS.tmp.1485W}, we will
test the three bracketing reionization histories first established in 
\citet{2015MNRAS.452..261C}, referred to as HM12, Late and Very Late. 
The evolution of the mass-averaged global neutral fraction
in these models can be seen in the left panel of Figure~\ref{fig:calib} with the solid lines.
We note that in the early
HM12 model, reionization ends (i.e. when $Q(z_\mathrm{end})=1$) at $z=6.7$. 
The Late model is the same as the HM12 model but shifted in redshift so that
reionization is completed by $z=6$. Finally in the Very Late model reionization 
also ends  at $z=6$ but with a different redshift gradient $\mathrm{d}Q/\mathrm{d}z$,
so that it finishes more abruptly. In this work we also consider modified versions of these
three models in which the end of reionization is delayed to $z\sim5.3$, but with
the same evolution at higher redshifts. These will be referred to as the
``Delayed'' models (in comparison to the ``Original'' models). Physically
these delayed models can result from an ionizing emissivity evolution that peaks
around $z\sim7$ and then falls dramatically at lower redshifts, as suggested
in \citet{2018arXiv180104931P}.
We note that the Delayed Very Late reionization history has a similar neutral
fraction evolution to the model
of \citet{2018arXiv180906374K}, which was found to reproduce the opacity
fluctuations in the Lyman-$\alpha$ forest (after reionization). The Delayed
models are shown in Figure~\ref{fig:calib} with the dash-dotted lines.
In the right panel of Figure~\ref{fig:calib} we compare the electron scattering
optical depths predicted by these histories with the recent Planck CMB 
measurements. The low value of $\tau = 0.054 \pm 0.007$ measured by
\citet{2018arXiv180706209P} favours a later reionization, and we see that our
Very Late models are consistent within 1$\sigma$ of this recent measurement.

These reionization histories are then calibrated on the simulation so that the
background photoionization rate, $\Gamma_\mathrm{HI}$, (which is naturally 
coupled to the global average ionization fraction) is self-consistent. The 
strength of the self-shielding and the equilibrium value of the neutral fraction
within ionized regions is dependent on this UV background.
Our calibrated
simulation is also consistent with observed background photoionization rates and 
ionizing photon mean free paths
\citep{2009ApJ...703.1416F,2011MNRAS.412.2543C,2011MNRAS.412.1926W} 
and the observed CMB optical depth
\citet{2018arXiv180706209P}. We note that our Very Late models are also consistent
with current observational estimates from the SILVERRUSH survey on the neutral 
fraction evolution (see appendix~\ref{appendix:Qtable}).


\begin{figure*}
    \includegraphics[width=\textwidth]{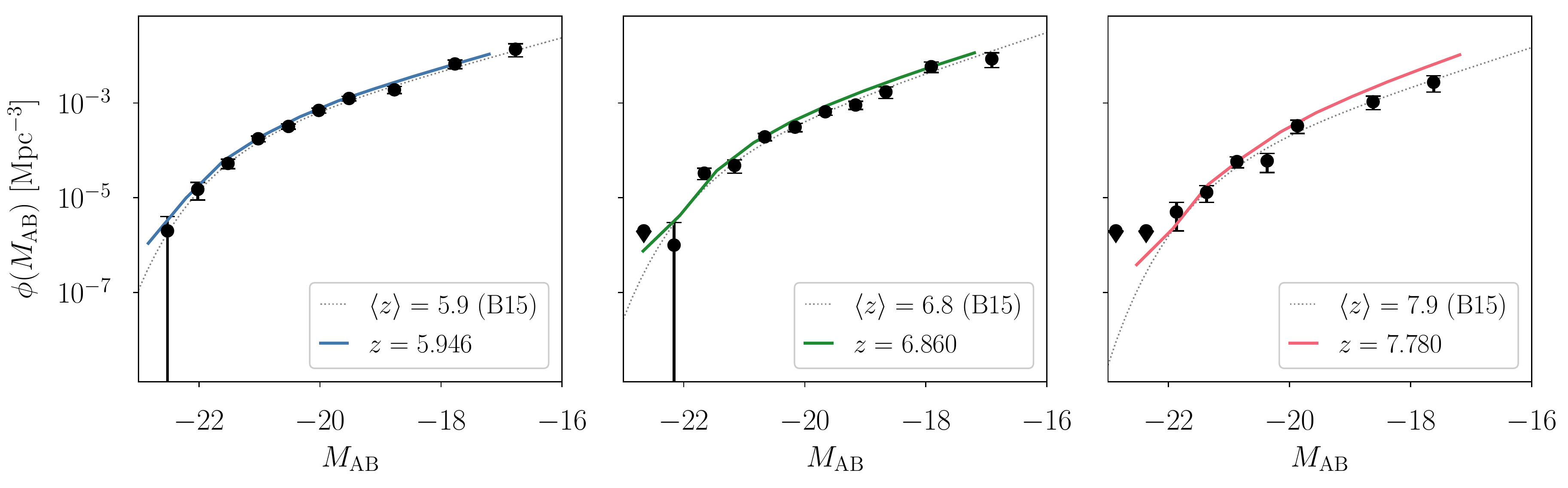}
    \caption{The evolution of the LBG UV luminosity function. The thick solid
     lines show our model predictions, compared to the data-points from 
     \citet{2015ApJ...803...34B} (and the dotted line best fit Schechter functions).
     The abundance
     matching is performed at $z=5.9$, so the agreement in the left-most panel
     is by construction. 
     At higher redshift the evolution is due to the evolution of the 
     halo mass function, and we see good agreement between our model predictions
     and the observed luminosity functions.}
    \label{fig:UVLF}
\end{figure*}

\subsection{Populating haloes with LBGs}
\label{sec:LBGs}
We implement the Improved Conditional Luminosity Function (ICLF) method of
\citet{2010ApJ...714L.202T} in order to populate the dark matter haloes in
our simulations with LBG-type galaxies. This involves abundance matching the
halo mass function to the observed UV luminosity function 
\citep[as in, e.g.][]{2009ApJ...695..368L, 2008ApJ...686..230B}. This assumes
that each halo hosts one galaxy, and then equates the number of halos
above a certian mass $M_h$ with the number of LBGs above a certain luminosity
$L_\mathrm{UV}$, 
\begin{equation}
    \epsilon_\mathrm{DC}(M_h,z) \int_{M_h}^\infty n(M,z) \dint M = 
    \int_{L_\mathrm{UV}}^\infty \phi(L,z) \dint L,
    \label{eq:abun}
\end{equation}
where $n(M,z)$ is the halo mass function, $\phi(L,z)$ is the UV luminosity
function, and $\epsilon_\mathrm{DC}(M_h,z) \leq 1$ is a mass and redshift dependent duty cycle which
accounts for how likely it is that we will observe the galaxy hosted by a given
halo at a particular time. Eq.~(\ref{eq:abun}) implicitly defines a mapping between halo
mass and UV luminosity, $L_\mathrm{UV}(M_h)$.
We use the same form of the duty cycle as \citet{2010ApJ...714L.202T},
\begin{equation}
    \epsilon_\mathrm{DC}(M_h,z) = 
    \frac{\int_{M_h}^\infty \dint M \left[n(M,z) - n(M,z_{\Delta t}) \right]}
         {\int_{M_h}^\infty \dint M \: n(M,z)}.
\end{equation}
Note that,
\begin{equation}
    \Delta t = t_H(z) - t_H(z_{\Delta t}),
\end{equation}
where $t_H$ is the local Hubble time \citep{2009ApJ...694..879T}. $\Delta t$ is 
a free parameter in this duty cycle model; in this work we choose $\Delta t=50$
Myr \cite[originally a duty cycle with $\Delta t=200$ Myr was employed by][]{2010ApJ...714L.202T}.
The time interval between $z=10$ and $z=6$ (when reionization is underway)
is less than 500 Myr, and so we choose a smaller $\Delta t$ to reflect the
bursty nature of star formation across this period. We note that other
numerical work has found similar time scales for luminosity variation between 
10--100 Myr, such as \citet{2018MNRAS.479..994R} who found large temporal
variation of the fraction of escaping ionizing photons, $f_\mathrm{esc}$, 
from galaxies during reionization.
Our choice will be further justified when we present the effect that varying
$\Delta t$ has on the clustering signal in section~\ref{sec:dt}. 
The $\Delta t=50$ Myr duty cycle as 
a function of redshift and mass is shown in Figure~\ref{fig:edc}. We finally note
that the $\Delta t$ parameter itself may in reality evolve with cosmic time, which 
would affect the evolution of the clustering signal of galaxies.

A heuristic picture of this duty cycle is as follows: although we expect galaxies
to exist in most dark matter haloes (above a minimum mass), it is not the case
that we expect these galaxies to always be bright enough (in the UV) to be observed.
The stochastic nature of star formation, and possibly geometric radiative transfer 
effects, might allow us to only observe a fraction of the underlying galaxy 
population (at a given time). In the
above abundance matching procedure, enforcing a duty cycle will alter the mapping
$L_\mathrm{UV}(M_h)$ in such a way as to shift the mapping to lower masses. This means
lower mass haloes can be brighter (compared to the result if $\epsilon_\mathrm{DC}=1$),
which will further impact the clustering. Recently, \citet{2018arXiv180700006G} explored the role
of radiative transfer effects within a given Ly $\alpha$ emitting galaxy, and
found that carefully including these effects for a population of LAEs causes the
LAEs to reside in less massive host haloes (compared to when such RT effects are
neglected). Although our duty cycle is somewhat
more agnostic to specific internal galaxy physics, it has the same effect. As in
that work, we find that populating less massive haloes leads to better agreement
with observables such as the clustering signal \citep{2009ApJ...695..368L}.

We perform the abundance matching detailed in Eq.~(\ref{eq:abun}) using the
Sheth-Mo-Tormen (SMT) halo mass function \citep{2001MNRAS.323....1S}, with \textsc{HMFcalc} 
\citep{2013A&C.....3...23M}. For the
UV luminosity functions, we employ the best-fit Schechter functions from
\citet{2015ApJ...803...34B}. In particular we calibrate our luminosity mapping
$L_\mathrm{UV}(M_h)$ at $\langle z\rangle=5.9$, and can then apply this to the
halo populations at each redshift of interest. \citet{2010ApJ...714L.202T} found
that their mapping, calibrated at $z\sim6$, was able to capture the evolution 
down to $z\sim4$ of the observed UV luminosity function. The shape of
$L_\mathrm{UV}(M_h)$ we find is similar to that found by \citet{2010ApJ...714L.202T},
but shifted to lower masses due to the lower $\Delta t$ that we employ in this work.

\begin{figure*}
    \fbox{\includegraphics[width=\textwidth]{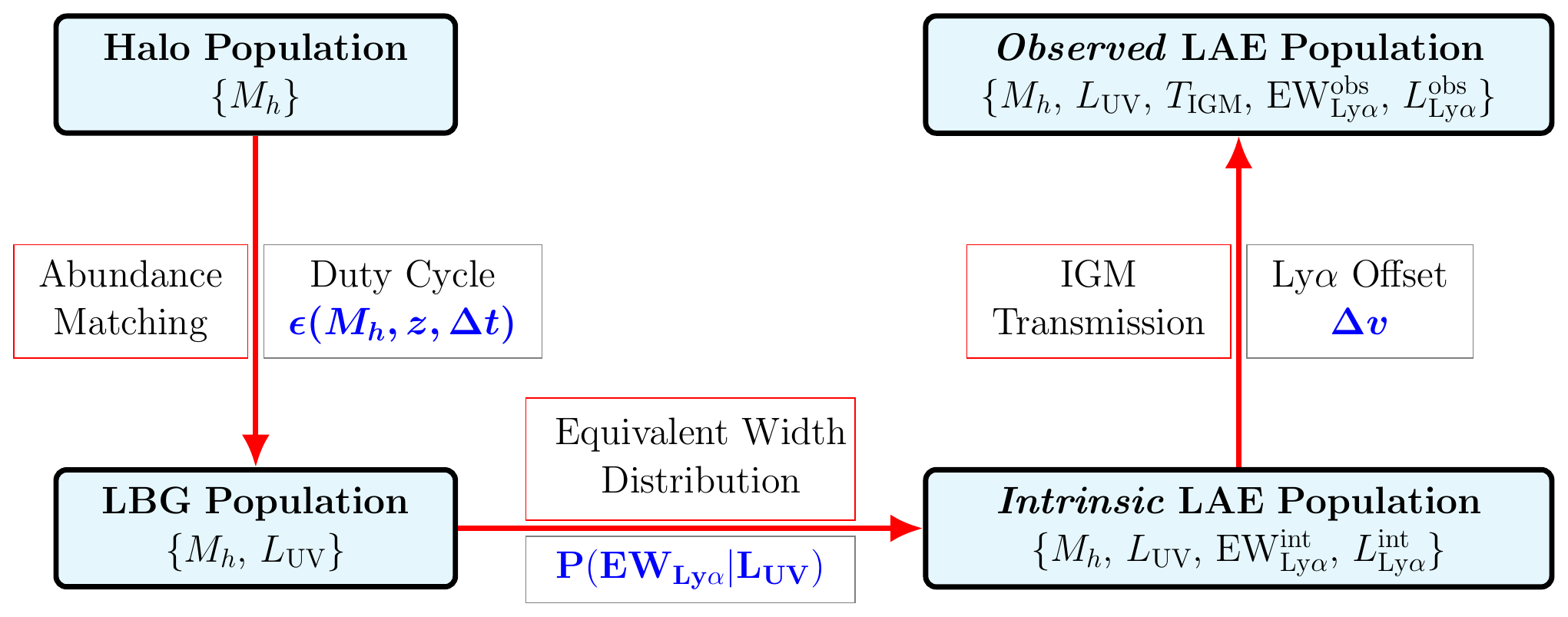}}
    \caption{Summary schematic representing the different parts of the LAE 
modelling: first the halo population of a given simulation snapshot is abundance
matched to the observed UV population with an assumed duty cycle, to create a
mock LBG population. Then we draw from an empirical `intrinsic' REW distribution
to create a mock `intrinsic' LAE population. Finally the effects of IGM transmission
are taken into account to create a mock `observed' LAE population.}
    \label{fig:schematic}
\end{figure*}

We can test that the evolution of the UV luminosity function is well fit by the 
underlying halo mass evolution by using the $L_\mathrm{UV}(M_h)$ mapping on
the simulated halo populations.
In Figure~\ref{fig:UVLF} we show this evolution starting
with the calibrated redshift $z=5.9$ on the left, and then higher redshifts
on the right. Note the average redshift of the observational sample can be
compared to the precise redshift of the simulation snapshot in the lower righthand
subplot legends.

After this step in the framework, we now have an observationally calibrated
mock sample of LBGs in our simulation. A subset of these will end up as a final
LAE mock sample, after the selection detailed below.


\subsection{LAE equivalent width distribution}
\label{sec:LAEs}
We implement the rest-frame equivalent width (REW) distribution model proposed
by \citet{2012MNRAS.419.3181D} in order to determine which of our LBG population
will have observable Ly $\alpha$ emission. This is calibrated to empirical
equivalent width distributions, therefore bypassing the need to model Ly $\alpha$ escape
fractions and emission mechanisms. The model starts with a $\muv$-dependent
distribution of Ly $\alpha$ REWs, 
\begin{equation}
    P(\rew|\muv) = \mathcal{N} \exp{\left(-\frac{\rew}{\rew_c(\muv)}\right)},
\end{equation}
where $\rew_c(\muv)$ is a charactestic REW for a given $\muv$,
given by \citep[the best fit model of][]{2012MNRAS.419.3181D},
\begin{equation}
    \rew_c(\muv) = 23 + 7(\muv + 21.9) + 6(z - 4).
\end{equation}
The normalization is defined such that the population has 
$\rew_\mathrm{min} \leq \rew \leq \rew_\mathrm{max}$, where we choose 
$\rew_\mathrm{max}=300$ \AA, and $\rew_\mathrm{min}$ is a function of $M_\mathrm{UV}$,
\begin{equation}
    \rew_\mathrm{min} = \\
    \begin{cases}
        -20 \text{ \AA} & \quad \muv < -21.5; \\
        17.5 \text{ \AA} & \quad \muv > -19.0; \\
        -20 + 6(\muv + 21.5)^2 \text{ \AA} & \quad \mathrm{otherwise}.
    \end{cases}
\end{equation}
Therefore the explicit form for the ($\muv$-dependent) normalization is,
\begin{equation}
    \mathcal{N} = \frac{1}{\rew_c} \left(\exp{\left(\frac{-\rew_\mathrm{min}}{\rew_c}\right)} 
    - \exp{\left(\frac{-\rew_\mathrm{max}}{\rew_c}\right)} \right)^{-1}
\end{equation}
This choice of $\muv$ dependence reproduces the so-called Ando 
relation \citep{2007PASJ...59..717A}.
We note \citet{2012MNRAS.419.3181D} compared this 
choice of distribution to the observed distributions at $z=$ 3.1, 3.7 and 5.7
by \citet{2008ApJS..176..301O}. They found good agreement at the lower redshifts,
but that their model over-estimated the number of large REW systems at the higher
redshifts. This was
calculated in the absence of attenuation by the CGM/IGM, which is non-zero
even after reionization has completed at $z\sim6$. We find that the inclusion
of an IGM transmission fraction $<1$ compensates for the overprediction, and hence
we do not modify the model to try to correct this. We further discuss this 
in section~\ref{sec:LF}.

\begin{figure*}
    \includegraphics[width=2\columnwidth]{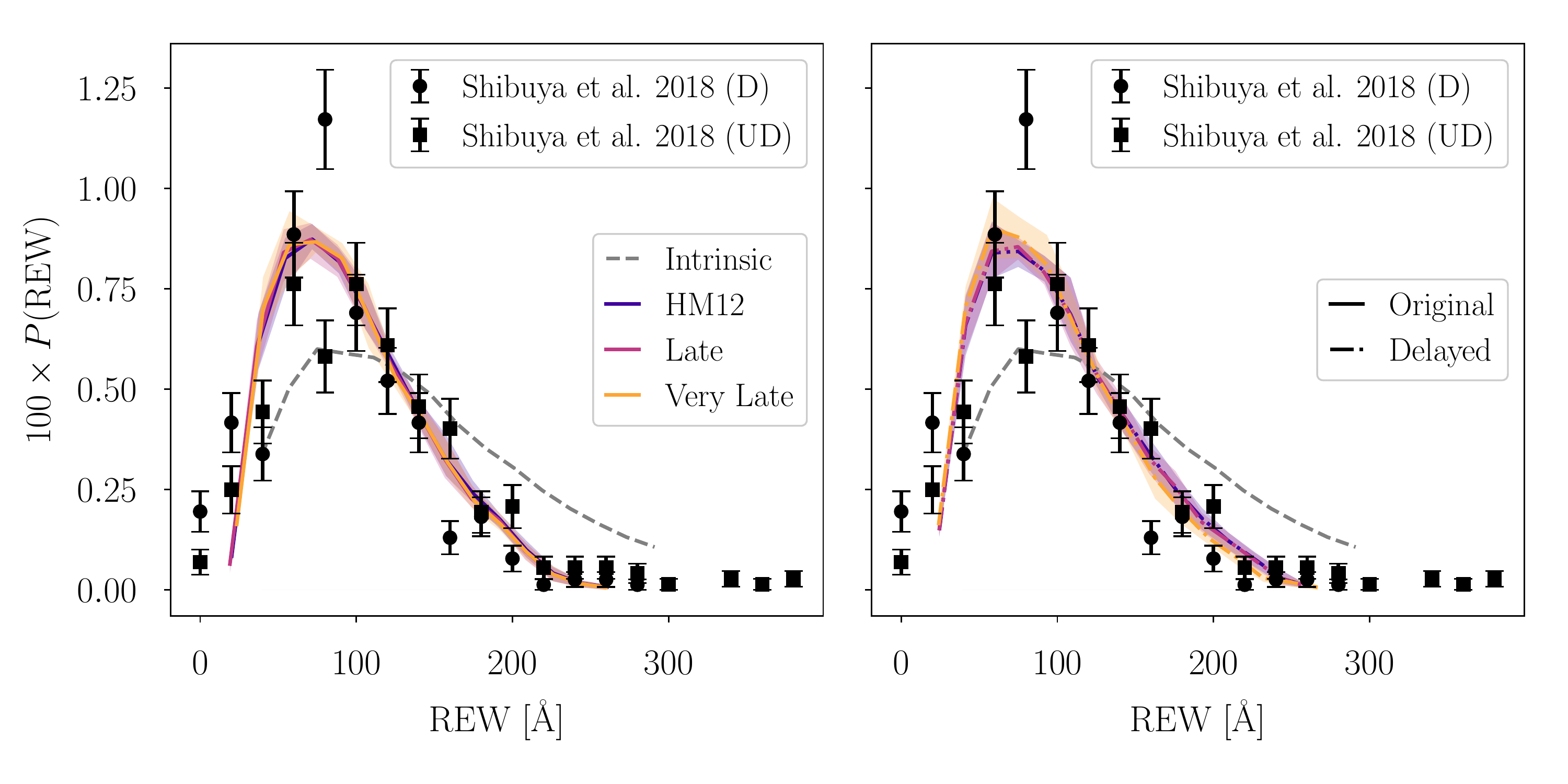}
    \caption{Comparison of our model mock REW distribution at $z=5.756$ with the observed
distribution from the SILVERRUSH survey at $z=5.7$. The different reionization
histories are shown by the different colours, whilst the intrinsic un-attenuated
model is shown with the grey dashed line. On the left hand panel in the solid
lines we show the original reionization histories; on the right hand panel we show
the Delayed models with dash-dotted lines. For the chosen emission offset 
parameter, $\Delta v \propto v_\mathrm{circ}$, we find that the IGM transmission
corrects the over-abundance of high REW systems predicted by the intrinsic model.}
    \label{fig:REW_dist}
\end{figure*}

For each of the haloes in our LBG sample, we draw a random REW from this 
conditional probability distribution, thereby assigning them a Ly $\alpha$
luminosity defined by,
\begin{equation}
    L_\mathrm{Ly\alpha} = \frac{\nu_\alpha}{\lambda_\alpha}
    \left(\frac{\lambda_\mathrm{UV}}{\lambda_\alpha}  \right)^{-\beta-2}
    \times \rew \times L_\mathrm{UV,\nu},
\end{equation}
where $\nu_\alpha=2.47\times 10^{15}$ Hz is the Ly $\alpha$ transition frequency,
$\lambda_\alpha=1216$ \AA{} the corresponding Ly $\alpha$ wavelength, 
$\lambda_\mathrm{UV}=1600$ \AA{} is the rest-frame wavelength at which the UV
luminosity function was measured \citep{2015ApJ...803...34B}, $\beta=-1.7$ is the
assumed UV spectral 
index, and the UV luminosity density, $L_\mathrm{UV,\nu}$, is related to $\muv$ by
\citep{2008ApJS..176..301O},
\begin{equation}
    M_\mathrm{UV} = -2.5\log{ L_\mathrm{UV,\nu} } + 51.6.
\end{equation}
Having generated $L_\mathrm{Ly\alpha}$ for the LBG mock sample, we then
apply selections based on luminosity and equivalent width limits to match a given
observational study. Table~(\ref{tab:thresholds}) shows some of the 
observational thresholds used in the SILVERRUSH survey 
\citep{2017arXiv170407455O,2017arXiv170501222K,2018PASJ...70S..14S}, 
which we adopt here. 

\begin{table}
\caption{Observational selection thresholds$^\dagger$ used in
this work to generate mock observed samples.}
\begin{tabular}{l|c|c|c}
\hline
Based on survey & $z$ & REW$_\mathrm{min}$ [\AA] & $L_\mathrm{Ly\alpha,min}^{\dagger\dagger}$ [erg/s] \\ \hline 
 \citet{2017arXiv170501222K} & 5.7 & 10 & 6.3$\times10^{42}$ \\ \hline
 \citet{2017arXiv170501222K} & 6.6 & 14 & 7.9$\times10^{42}$ \\ \hline 
 \citet{2017arXiv170302501O} & 7.0 & 10 & 2$\times10^{42}$ \\
 \citet{2018arXiv180505944I} &     &    &                  \\ \hline
 \citet{2014ApJ...797...16K} & 7.3 & 0  & 2.4$\times10^{42}$ \\ \hline
\end{tabular}
\vspace*{0.1cm}

{\footnotesize \textbf{Note}:

$^{\dagger}$The observational surveys we derived these limits from 
measured different fields on the sky, across which
different selection thresholds were sometimes applied. The values we have chosen
are representative despite this variation.

$^{\dagger\dagger}$As in \citet{2012MNRAS.419.3181D}, when not quoted in the 
original survey, we estimate $L_\mathrm{min}$
as the lower bin edge of the lowest luminosity bin in the presented luminosity
function.}
\label{tab:thresholds}
\end{table}

We note that
\citet{2012MNRAS.419.3181D} concluded that equivalent width and luminosity cuts
are only an approximation to the real selection thresholds used in observational
LAE studies. They found that in order to match both the observed equivalent width
distribution and the luminosity function, they had to scale the number density
by $\sim0.5$. With this scaling included, their modelling then matched the redshift
evolution very well.

When calculating the IGM transmission (as described in detail in section 
~\ref{sec:transmission}) in our simulations we
find that even at $z=5.7$ the transmission redwards of Ly $\alpha$ is not 100\%.
As was suggested in \citet{2017ApJ...839...44S}
\citep[and explored further in][]{2018MNRAS.tmp.1485W}, 
this is due mostly to the neutral gas in the outer part of the host halo (close
to the virial radius),
controlled by the background photoionization rate (see section~\ref{sec:CGM}). 
In our delayed models there is 
also an increased fraction of neutral gas in the rest of the IGM, but we find the
attenuation at these redshifts is dominated by CGM and surrounding halo gas
\footnote{In this work we will refer to the infalling gas surrounding the 
halo as the CGM gas. We note however that we include gas out to larger scales
($r \lesssim 10 R_\mathrm{vir}$) than the more common observational definition of the
CGM gas \citep[at scales closer to $R_\mathrm{vir}$, e.g.][]{2010ApJ...717..289S}}.
This means that when we include the transmission we find the number density of
`observed' LAEs drops, in good agreement with the real observed 
luminosity functions. We therefore choose to use the REW distribution from
\citet{2012MNRAS.419.3181D} to model the `intrinsinc' distribution of our 
population, which is then attenuated by the IGM to give the `observed' REWs and
Ly $\alpha$ luminosities. As such we do not apply the $\sim0.5$ scaling 
to the number density in our model predictions which was needed in 
\citet{2012MNRAS.419.3181D}; we find agreement with the observational results
without this.

\begin{figure*}
    \includegraphics[width=2\columnwidth]{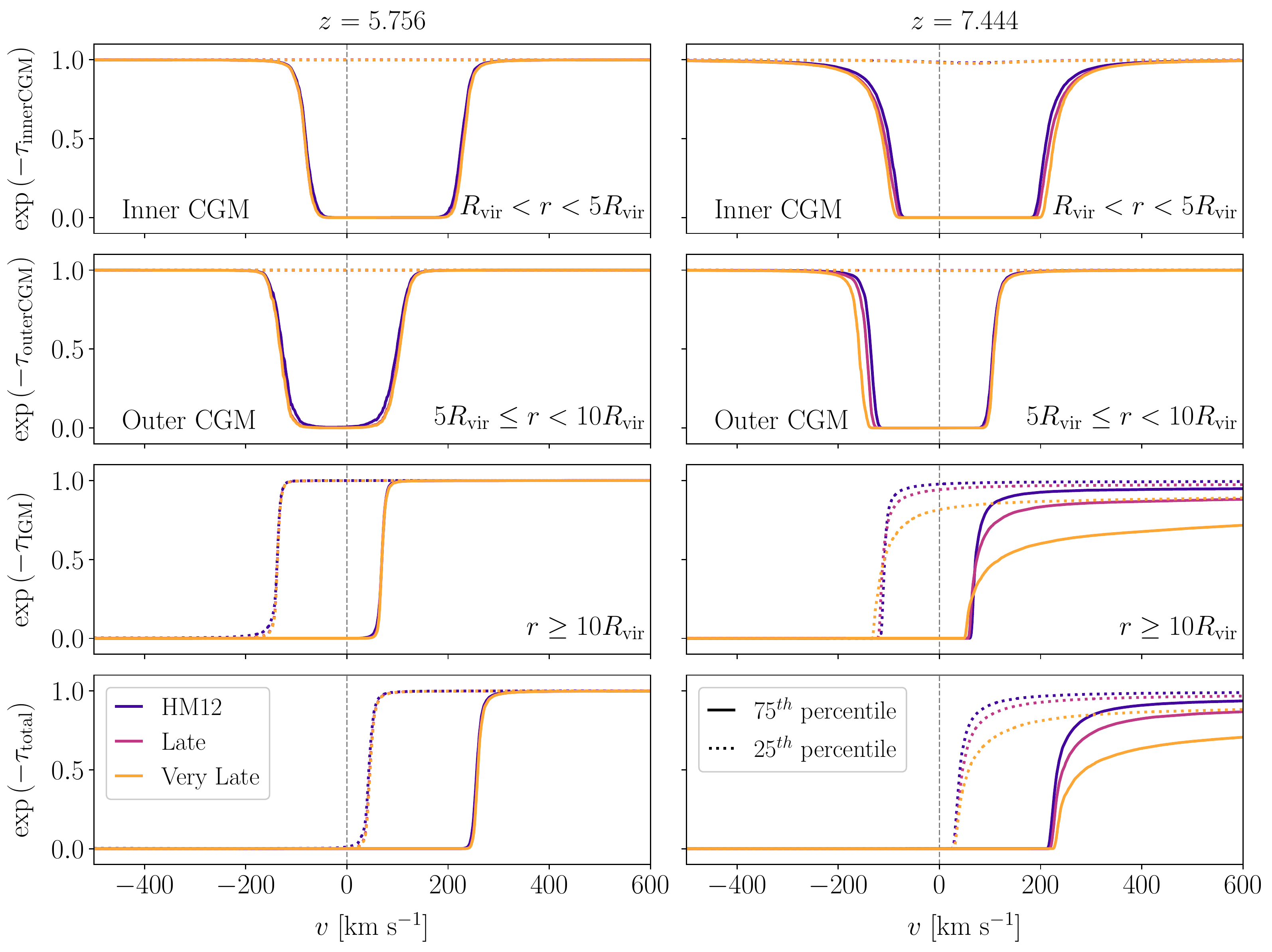}
    \caption{Separating the contributions to the Ly $\alpha$ transmission: (i) in the top panels 
    we show the  transmission due to the CGM and surrounding gas (split into
    an ``inner'' part between $R_\mathrm{vir} < r < 5 R_\mathrm{vir}$ and an
    ``outer'' part between $5 R_\mathrm{vir} \leq r < 10 R_\mathrm{vir}$)
    (ii) in the middle panels the contribution from the IGM
    ($r \geq 10 R_\mathrm{vir}$); 
    (iii) in the bottom panels the transmission from all the gas outside the halo ($r > R_\mathrm{vir}$). 
    In all panels the solid (dotted) curves correpond to the $75^{th}$ 
    ($25^{th}$) percentile as measured across our mock observed sample of
    LAEs (i.e. after $L_\mathrm{Ly\alpha}$ and REW selection), which spans an
    order of magnitude in luminosity.
    The transmission is shown
    as a function of velocity offset from line-centre, at redshift $z=5.756$ on the
    left and $z=7.444$ on the right. The three original reionization histories 
    are shown using the coloured solid lines. For clarity we do not plot 
    the delayed-end models, however their corresponding transmission curves are
    very similar. }
    \label{fig:components}
\end{figure*}

This means we now have a mock LAE sample, in which each
object has a value of $M_h$, $M_\mathrm{UV}$, REW, and
$L_\mathrm{Ly\alpha}$ that conforms to the chosen observational selection
window. The final step in the framework, before we can compare with observations,
is to calculate the IGM transmission fraction for a given halo in the sample, and reduce
the luminosity accordingly.

\begin{figure*}
    \includegraphics[width=\textwidth]{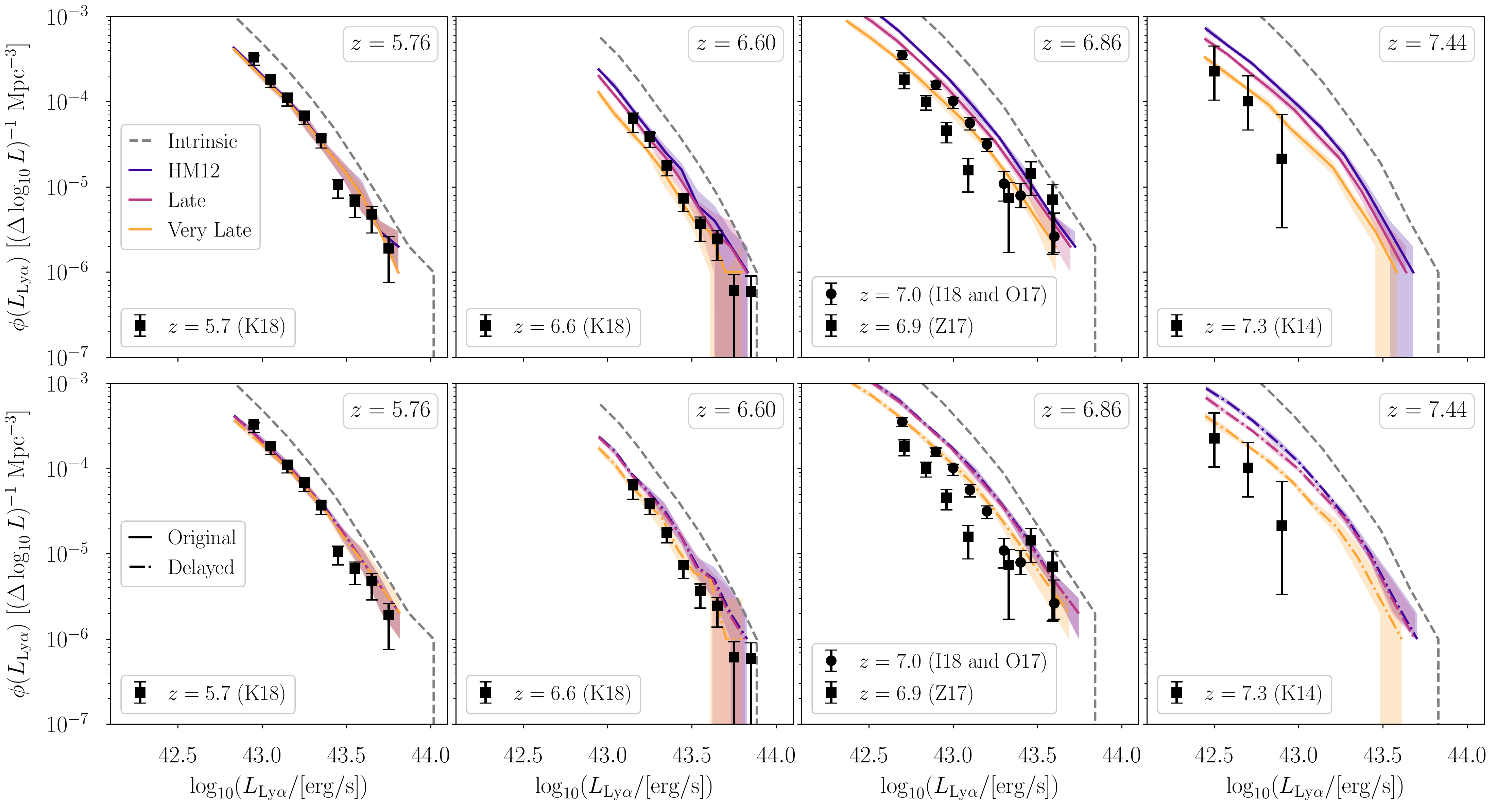}
    \caption{From left to right: The luminosity function evolution,
     at $z=5.756$, 6.604, 6.860 and 7.444, for each of the
    reionization histories shown as coloured lines. The top panel with solid
    lines shows the original three reionization histories, whilst the bottom
    panel with dash-dotted lines shows the delayed-end histories. 
    Some observed data
    is overplotted for comparison with black markers, from 
    \citet{2017arXiv170501222K} (K18, $z=5.7$ \& 6.6),
    \citet{2018arXiv180505944I} (I18, $z=7.0$), 
    \citet{2017arXiv170302501O} (O17, $z=7.0$), 
    \citet{2017arXiv170302985Z} (Z17, $z=6.9$), 
    \citet{2014ApJ...797...16K} (K14, $z=7.3$).} 
    \label{fig:LF}
\end{figure*}


\subsection{Ly $\alpha$ transmission}
\label{sec:transmission}
We calculate the CGM/IGM transmission as in \citet{2018MNRAS.tmp.1485W}, extracting
sightlines\footnote{This includes a 20 cMpc/h region around the halo with a higher 
resolution of 9.8 ckpc/h.} through the halo sample to find the optical depth to Ly $\alpha$, 
$\tau_\mathrm{Ly\alpha}(v)$, as a function of velocity offset from the emitter
\citep{2007MNRAS.374..493B}.
This can be used to calculate a transmission fraction, assuming an intrinsic
emission profile $J(v)$ for the galaxy, which itself accounts for the radiative transfer
within the halo as photons escape the galaxy's ISM. The emission profiles of LAEs
seen in both observations \citep[e.g.][]{2015ApJ...812..157H} and 
radiative transfer simulations 
\citep[e.g.][]{2014ApJ...794..116Z} are non-trivial to model, 
with complicated dependences on the local gas dynamics. 

We make the simplifying 
assumption that the emission profile is a single-peaked
Gaussian profile with width $\sigma_v$, offset redwards from the 
systemic by $\Delta v$ due to resonant
scattering within the halo. We choose $\sigma_v = 88$ 
km/s as in \citet{2015MNRAS.452..261C}. Our fiducial choice for the 
velocity offset is to set it proportional to the 
virial circular velocity of a given LAE's host halo, $\Delta v \propto v_\mathrm{circ}$.
This is motivated by observational and theoretical work such as 
\citet{2018MNRAS.478L..60V,2018arXiv181008185S,2006ApJ...649...14D,1990ApJ...350..216N},
which has shown that the radiative transfer (as a strong function of HI opacity)
 in a galaxy's ISM/CGM leads to a coupling
 between the dispersion of the Ly $\alpha$ line and its velocity
offset. We might expect the dispersion to be proportional to a halo's circular
velocity, which in turn gives the same proportionality for the offset. 
Previously in 
\citet{2018MNRAS.tmp.1485W} we used a fixed value of $\Delta v = 100$ km/s,
neglecting any dependence on the emitter properties and assuming no variation
across the population. Although this choice was simplistic, we found that 
the relative transmission fraction, 
$T^\mathrm{IGM}_\mathrm{Ly\alpha}(z)/T^\mathrm{IGM}_\mathrm{Ly\alpha}(z=5.7)$,
was largely insensitive to this choice. In the present work we now need to 
consider the absolute transmission fraction at a given redshift and so we have 
updated our model for the intrinsic emission.

\begin{figure*}
    \includegraphics[width=\textwidth]{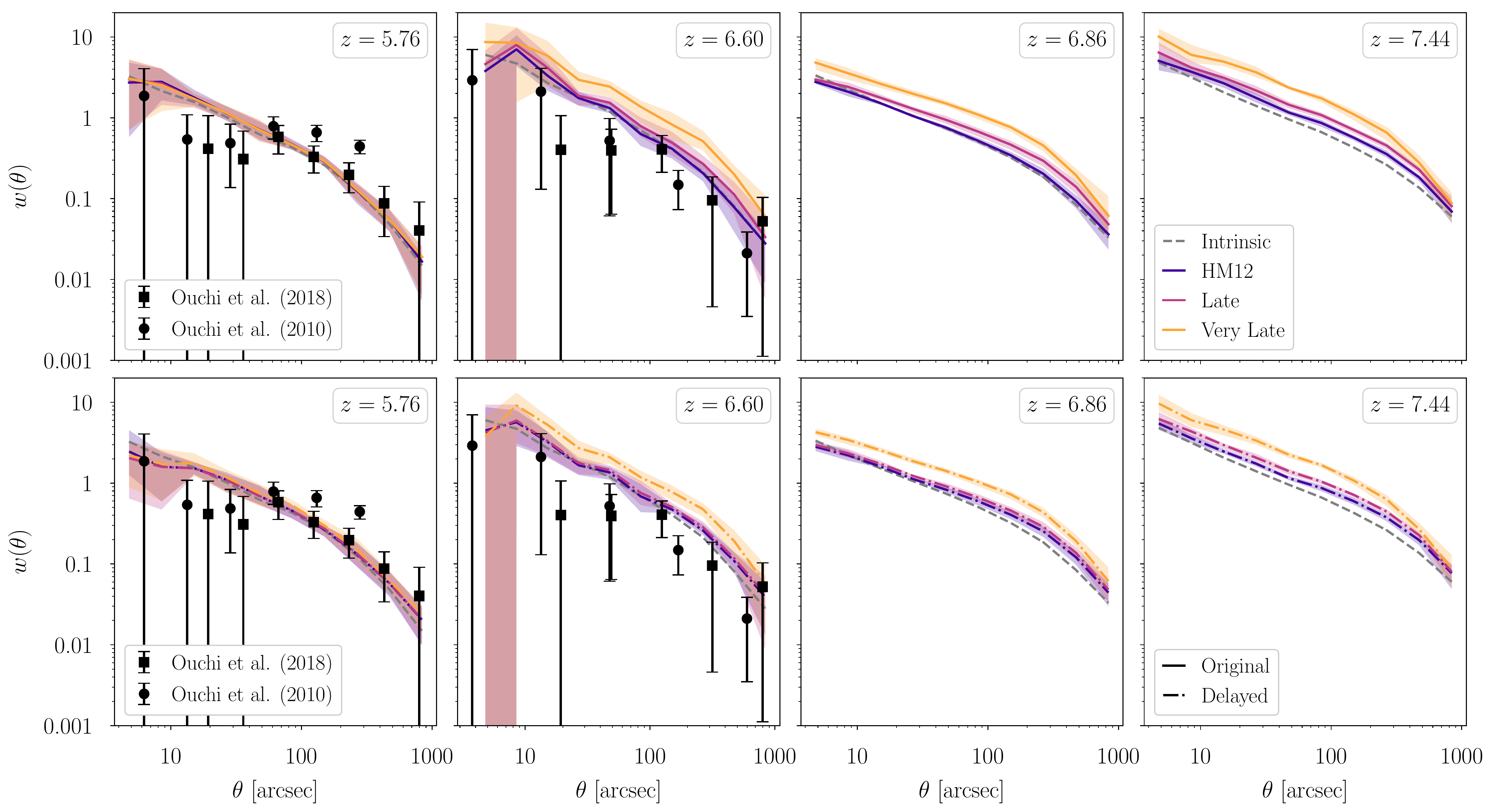}
    \caption{From left to right: The angular correlation function, measuring
    the clustering signal at $z=5.756$, 6.604, 6.860 and 7.444, for our 
    fiducial model with $\Delta t= 50$ Myr. As in Figure 6, the top panel
    shows the original reionization histories in solid lines, whilst the bottom
    panel shows the delayed-end versions in dash-dotted lines.
    At $z=5.756$ the reionization histories have
    converged (reionization has ended) and so the models are equivalent.
    At higher redshifts we start to see some divergence of the
    amount of clustering predicted for each reionization history, with the Very 
    Late model predicting the most clustering power. For each redshift
    these mock surveys correspond to 10 slices of our simulation volume (each 
    with area $320\times320$ (cMpc/h)$^2$), therefore simulating
    a total survey area of $2.2$ Gpc$^2$. We also refer the reader to 
    Table~(\ref{tab:thresholds}) for the mock selection limits used at each 
    redshift.}
    \label{fig:acf}
\end{figure*}

Other theoretical works have treated the intrinsic emission profile differently,
for example \citet{2018arXiv180100067I} used the single peaked profiles that result from full 
radiative transfer calculations applied to an outflowing spherical volume of gas. 
This is strongly dependent on the chosen HI column density and outflow velocity; our 
modelling assumes an explicit halo mass dependence for the profile rather than
fixing these quantities across the population.

We finally note that this is a poorly constrained quantity observationally, 
and so we later test the effect of varying our $\Delta v$ assumption. 
In reality the intrinsic emission profile of an LAE will evolve 
and vary across its lifetime, a feature we do not attempt to model in this work.

The transmission
fraction for this profile can then calculated as \citep{2015MNRAS.446..566M},
\begin{equation}
    \label{eq:transmissivity}
    T^\mathrm{IGM}_\mathrm{Ly\alpha} = \frac
    {\int \dint \nu \: J(\nu) \: e^{-\tau(\nu)}}
    {\int \dint \nu \: J(\nu)}.
\end{equation}
Note that specifically this is the transmission fraction redwards of systemic;
in choosing a single-peaked profile we have accepted that the IGM is sufficiently
optically thick at the redshifts we consider such that even if a blue peak emerges after
radiative transfer in the ISM, the HI damping wing of the IGM will not
transmit bluewards of systemic. This method of separating the radiative transfer into
a galaxy stage (which we simply model with our velocity offset $\Delta v$)
and an IGM stage (which we model using the $e^{-\tau}$ approximation) has been
employed successfully before, for example in \citet{2018arXiv180607392L,
2011ApJ...728...52L}. As we account for the galactic radiative transfer already,
we exclude the host halo gas in our calculation of $\tau_\mathrm{Ly\alpha}$.
Our fiducial choice is to exclude halo gas within 1 $R_\mathrm{vir}$ of the 
halo centre, the effect of which was
tested in \citet{2018MNRAS.tmp.1485W}. Similarly \citet{2018arXiv180607392L}
chose to transition between the two calculation regimes at 1.5 $R_\mathrm{vir}$.
We reiterate that the choice of intrinsic emission profile has a strong
effect on the transmission fraction, and therefore also on the resulting population
statistics for our LAE mock catalogues.

Having performed the transmission fraction calculation for each halo, we can
update the derived Ly $\alpha$ luminosity,
\begin{equation}
    L^\mathrm{obs}_\mathrm{Ly\alpha} = T^\mathrm{IGM}_\mathrm{Ly\alpha}L_\mathrm{Ly\alpha},
    \label{eq:obsL}
\end{equation}
and similarly for the equivalent widths.
This completes our generation of a mock LAE sample from our simulated halo population.
This framework allows us to generate samples at any desired redshift to compare with
observations, and make predictions at higher redshifts for the evolution of the
luminosity function and clustering signal. Note that when the transmission
fraction falls below unity, some of the LAEs will drop below the flux-limit, 
hence we reapply the selection after accounting for transmission. 
In Figure~\ref{fig:schematic} we show a summary schematic for the different 
stages of the LAE modelling.


\subsection{Modelling caveats}
Although  this model framework is successful at matching the evolution of
observed luminosity functions and clustering studies 
(as we will present in section~\ref{sec:results}), there are a number of 
important caveats to consider:
\begin{itemize}
    \item We do not use sub-find catalogs with satellite haloes. As we are only 
using central haloes we will naturally underestimate the 1-halo term of the clustering
signal. However at the redshifts of interest $z\gtrsim6$ the halo occupation 
distribution (HOD) populates only very large mass haloes with satellites. For
example \citet{2018arXiv180610612B} found that the satellite fraction of haloes
is around $\sim10\%$ at $z=7.5$, and that the mean number of satellites is less
than unity for halo masses below $M_h\sim10^{11}\:M_\odot/h$. Similarly
\citet{2018arXiv180700006G} found that satellite galaxies only start to
dominate the abundance of haloes with mass $M_h \gtrsim 10^{12}\:M_\odot/h$.
    \item There are a number of tunable parameters in this model. The primary free
parameters are $\Delta t$ which controls the duty cycle, and $\Delta v$ which
controls the `intrinsic' emission profile and hence has a strong effect on the
calculated transmission fraction. We take the best-fit REW distribution from
\citet{2012MNRAS.419.3181D} and so do not leave any of those model parameters
free, but use their empirically constrained values.
    \item Although we predict the transmission, and leave the duty cycle as a free
parameter, these variables have a degenerate effect on the clustering. Increasing
(decreasing) the duty cycle (the transmission) can lead to an increase in the
measured clustering signal. We have used physically motivated values for the
free parameters $\Delta t$ and $\Delta v$.
\end{itemize}


\section{Results}
\label{sec:results}
We now discuss the results of applying our LAE framework to the halo population
in the Sherwood simulations. In particular we consider the IGM transmission
across this population, and how this affects the luminosity function evolution.
We also confirm that our model matches the observed equivalent width distribution
and luminosity function at $z=5.7$. Taking this redshift as an anchor we then
also create mock survey slices from which we calculate the angular correlation
function, a 2-point measure of the clustering signal. Finally we extend our
models to higher redshifts to make predictions for future surveys.

Note for all plots where we compare our model predictions to the observational 
data, we have taken slices of the simulation volume and calculated the
relevant statistical quantity for the sample of LAEs in each slice. The 
shading shown for the model predictions corresponds to 68\% scatter across the
slices. Specifically we divide the box up in configuration-space into 10 slices
perpendicular to the direction along which we calculated the transmission,
giving a comoving thickness of 32 cMpc/h. This is not exactly equivalent to the
narrowband selection, which is instead a slice in redshift (velocity) space;
however given the width of the narrowband slice and the comparatively small 
amplitudes of peculiar motions of the LAEs in our simulation volume, 
we find that the results using configuration-space slicing
are indistinguishable from the velocity-space slicing. In all figures we
show the intrinsic (unattenuated) quantities using a dashed grey line, whilst
the different model (attenuated) quantities are shown using coloured solid
lines (original reionization histories) and dash-dotted lines (delayed-end
models).

\subsection{IGM attenuation of the $z=5.7$ equivalent width distribution}
\label{sec:obsREW}
In Figure~\ref{fig:REW_dist} we compare the observed REW distribution from
the SILVERRUSH survey \citep{2018PASJ...70S..14S} at $z=5.7$ with that predicted
from our mock LAE population. We show the un-attenuated `intrinsic' distribution
with the dashed grey line, whilst the IGM attenuated distributions are shown
for the three different reionization histories in blue (HM12), purple (Late) and
orange (Very Late).

\begin{figure}
    \includegraphics[width=\columnwidth]{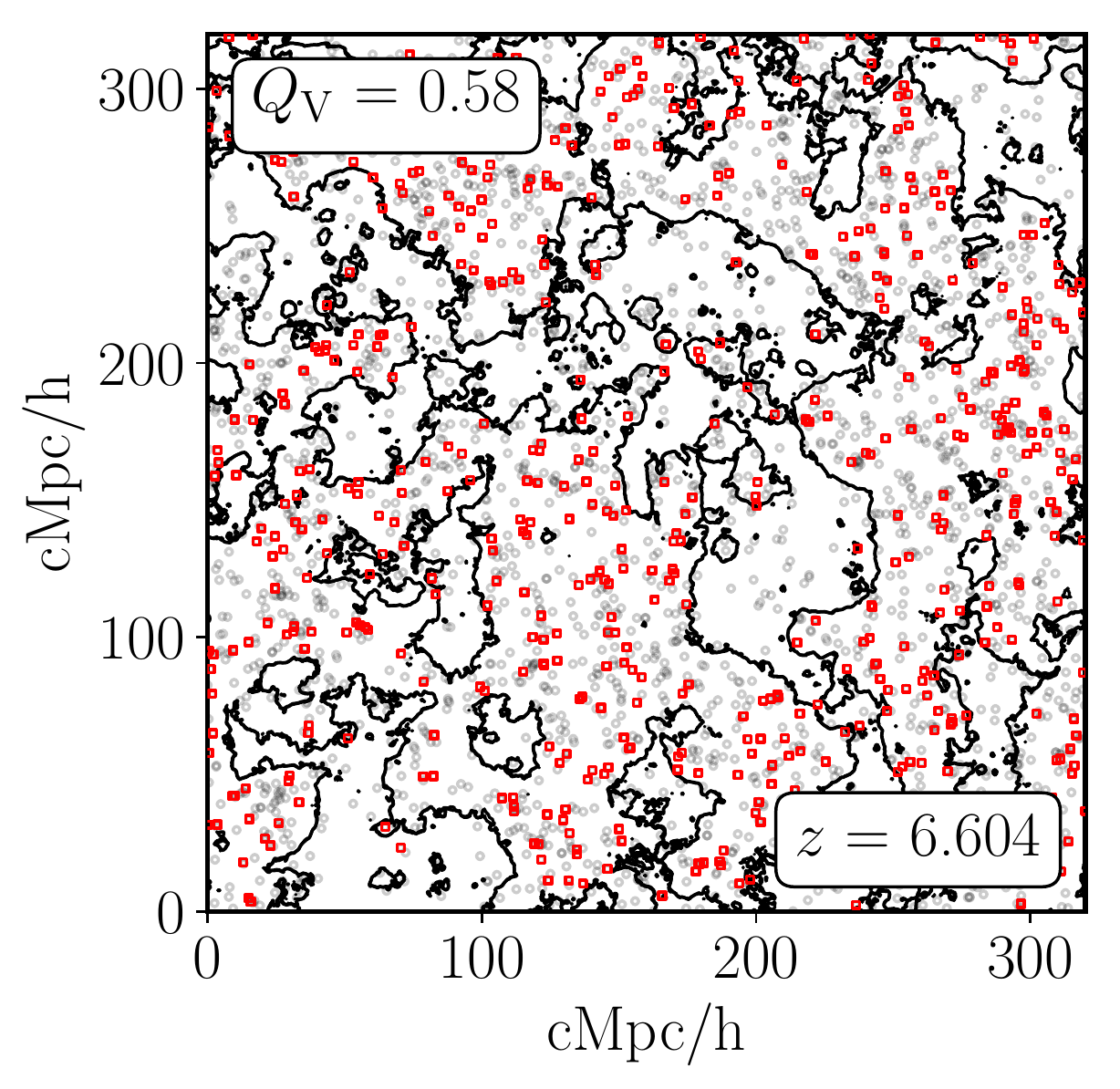}
    \caption{A mock map of a sample of LAEs within a 32 cMpc/h slice of the simulation
    volume, showing the whole population in grey empty circles and those above 
    the observational thresholds in red empty squares. 
    A black contour at a value of half the maximum projected neutral fraction is 
    also plotted, segregating the regions within this projected
    slice which are largely neutral from those which are ionized. It is visible
    by eye that the majority of the observed mock LAEs (in red) lie within
    the ionized regions.}
    \label{fig:map}
\end{figure}

We see at this redshift that although our original reionization histories give a global
average ionized fraction of unity, the IGM transmission fraction is sufficiently 
below unity that it has a significant effect on the observed REW distribution.
This attenuation results from (photoionization) equilibrium and self-shielded 
neutral gas around haloes which contributes little to the average ionized 
fraction but has a strong effect on the Ly $\alpha$ attenuation (see section
~\ref{sec:CGM}).
In particular the transmission fraction distribution is such that it attenuates
the high REW objects, thereby reducing the over-abundance of such objects which
is predicted by the intrinsic distribution. We have chosen to use a velocity 
offset of $\Delta v = a\: v_\mathrm{circ}$ where,
\begin{equation}
    \label{eq:avals}
    a =
    \begin{cases}
        1.5 & \mathrm{Original}, \\
        1.8 & \mathrm{Delayed},
    \end{cases}
\end{equation}
which gives a distribution consistent with the observed data. 
We also tested proportionality constants $a = 1$ and $a = 2$ and found that
these resulted in either too much attenuation or too little, respectively.
The larger value of $a$ is required in the Delayed models, where reionization
has not yet ended by $z=5.7$, so there is considerably more attenuation by
residual neutral gas. In particular we find that these models require a lower
background photoionization rate, therefore increasing the amount of neutral
gas present in the outer parts of the LAE host haloes (in the CGM). The presence of this
gas is sufficient to reduce the transmission redwards of Ly $\alpha$.

The resonant scattering of Ly $\alpha$ radiation by neutral hydrogen within
the galaxy tends to diffuse the emission profile away from line-centre, with radiation
escaping in blue or red peaks where the scattering cross-section is smaller. If
the galaxy has outflows, the red peak can be enhanced such that the dominant
emission comes at redder velocities. This is seen in shell models with an
expanding \ion{H}{I} outflow \citep[e.g.][]{2011MNRAS.414.2139D,2006A&A...460..397V}.
We might therefore expect that the velocity offset $\Delta v$ should be coupled to the
galaxy wind velocity which, to avoid stalling, must be of order the escape velocity
of the halo, i.e. $\sim \sqrt{2} v_\mathrm{circ}$. Hence the values of $a$
in Eq.~(\ref{eq:avals}) are reasonable.

\begin{figure}
    \includegraphics[width=\columnwidth]{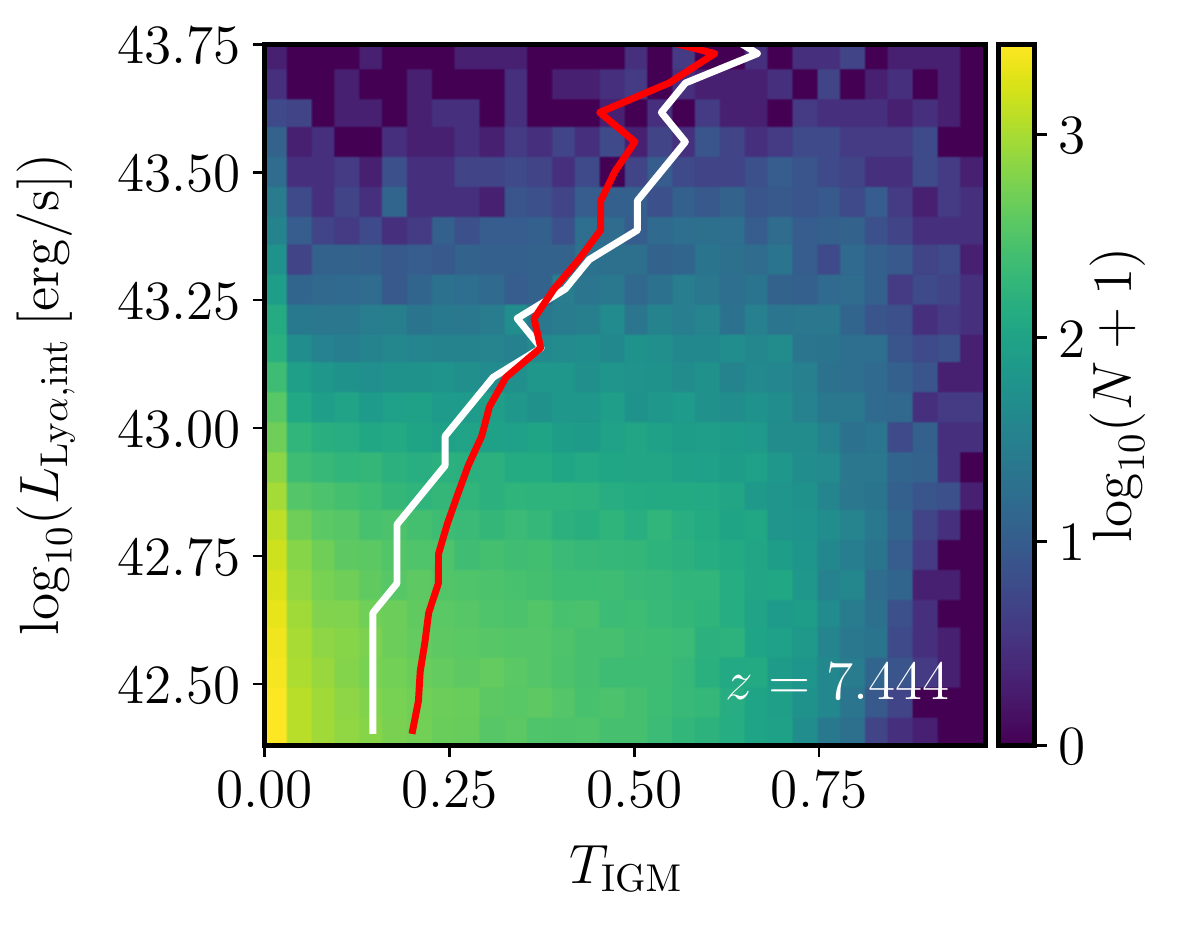}
    \caption{Histogram for the transmission fraction distribution and the
    `intrinsic' Ly $\alpha$ luminosity for our mock LAE population at $z=7.444$.
    The number of objects in each bin is indicated by the colourmap, whilst the
    red (white) line indicates the mean (median) transmission fraction for a 
    given luminosity bin. 
    We see that the brighter LAEs are preferentially more visible than the
    fainter LAEs.}
    \label{fig:brightfaint}
\end{figure}

\subsection{Attenuation from the partially neutral CGM}
\label{sec:CGM}
As seen in section~\ref{sec:obsREW}, even at $z=5.756$ (where our original
reionization histories are fully reionized, i.e. $Q_\mathrm{M}=1$) there is
still an attenuation of radiation redwards of Ly $\alpha$. This is due to
infalling neutral gas around the halo (including the CGM around the LAEs),
which is not fully ionized by either the LAE itself or the ionizing UV
background. \citet{2017ApJ...839...44S} found that the drop in observed numbers 
of LAEs doesn't
necessarily imply a largely neutral IGM, since this infalling CGM gas can
also bring the Ly $\alpha$ transmission below $100\%$. We note again that
our usage of the term CGM refers to a larger volume of the infalling gas
that surrounds the host halo than the more common observational definition.

To quantify this further, in Figure~\ref{fig:components} we plot the $75^{th}$
and $25^{th}$ percentiles of the transmission along sightlines to our observed 
samples of mock LAEs. We show the
transmission as a function of velocity offset from line-centre due to: (i) the
infalling CGM and surrounding gas (split into an ``inner'' part between 
$R_\mathrm{vir} < r < 5 R_\mathrm{vir}$ and an ``outer'' part between 
$5 R_\mathrm{vir} \leq r < 10 R_\mathrm{vir}$ around the halo 
center); (ii) the exterior IGM gas ($r \geq 10 R_\mathrm{vir}$); and (iii) the
total gas around the LAE.
We show this for our three original reionization histories
at the two bracketing redshifts\footnote{The transmission curves for the 
delayed-end models are similar, but are not shown to aid the clarity of Figure
\ref{fig:components}.}. We see that there is significant halo-to-halo variation
in the CGM component, where the transmission can vary from 100\% at all velocities
(in the $25^{th}$ percentile) to 0\% around $v=0$ km/s (in the $75^{th}$ percentile).
In the left panel (at $z=5.756$) we see that 
the IGM is transmitting at $\sim100\%$ redwards of Ly $\alpha$ (i.e. $v>0$ km/s) for 
the $25^{th}$ percentile, whereas the $75^{th}$ percentile starts to transmit at 
$\sim100\%$ for redder velocities $v\gtrsim100$ km/s.
Furthermore we see the drop in transmission due to the CGM gas
extends redwards of line-centre, because the gas is infalling onto the halo.
In particular the ``inner'' part of the CGM attenuates redwards of the ``outer'' part because
the amplitude of the infalling gas velocity peaks in that region.
This means that radiation redwards of line-centre can be blue-shifted in the
frame of the gas towards line-centre, and hence resonantly scattered out of the line of sight.
The CGM transmission evolves across the redshifts as a function of the
photoionization rate, which controls how neutral the gas is. 

\begin{figure}
    \includegraphics[width=\columnwidth]{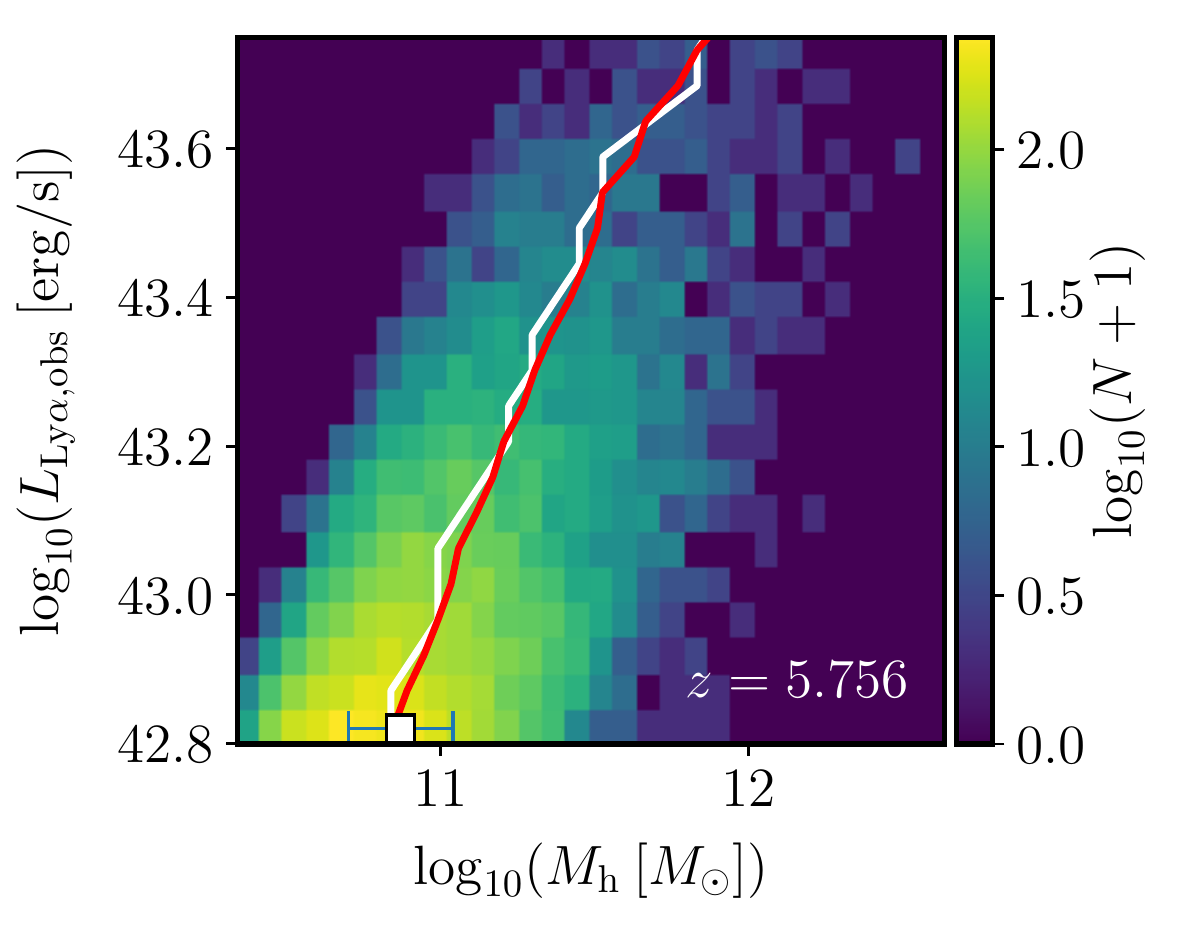}
    \caption{Histogram for the mass distribution and the
    `observed' Ly $\alpha$ luminosity for our mock LAE population at $z=5.756$.
    The number of objects in each bin is indicated by the colourmap, whilst the
    red (white) line indicates the mean (median) host halo mass for a 
    given luminosity bin. 
    The white datapoint shows the median luminosity and effective halo mass
    for the NB816 sample of \citet{2018arXiv181100556K}.}
    \label{fig:massdist}
\end{figure}

In comparison the IGM transmission gradually decreases with increasing redshift as the
average neutral fraction increases. Considering the shape of the attenuation
imprinted by these different components, we note that the
CGM evolution (dependent on the photoionization rate) causes a velocity shift
in the transmission curve along the horizontal axis, whilst the
IGM evolution (a function of the average neutral fraction) causes less transmission,
i.e. a shift along the vertical axis. In particular we note that near the end of
reionization the horizontal shift caused by the CGM is the dominant component
of the attenuation. At higher redshifts we can distinguish the different
reionization histories because their average neutral fractions diverge 
significantly, causing varying amounts of vertical shift in the transmission
curve. We find there is a luminosity (or mass) dependence in the evolution
of these two components, explored further in appendix~\ref{appendix:compareCGM}.

\begin{figure*}
    \includegraphics[width=2\columnwidth]{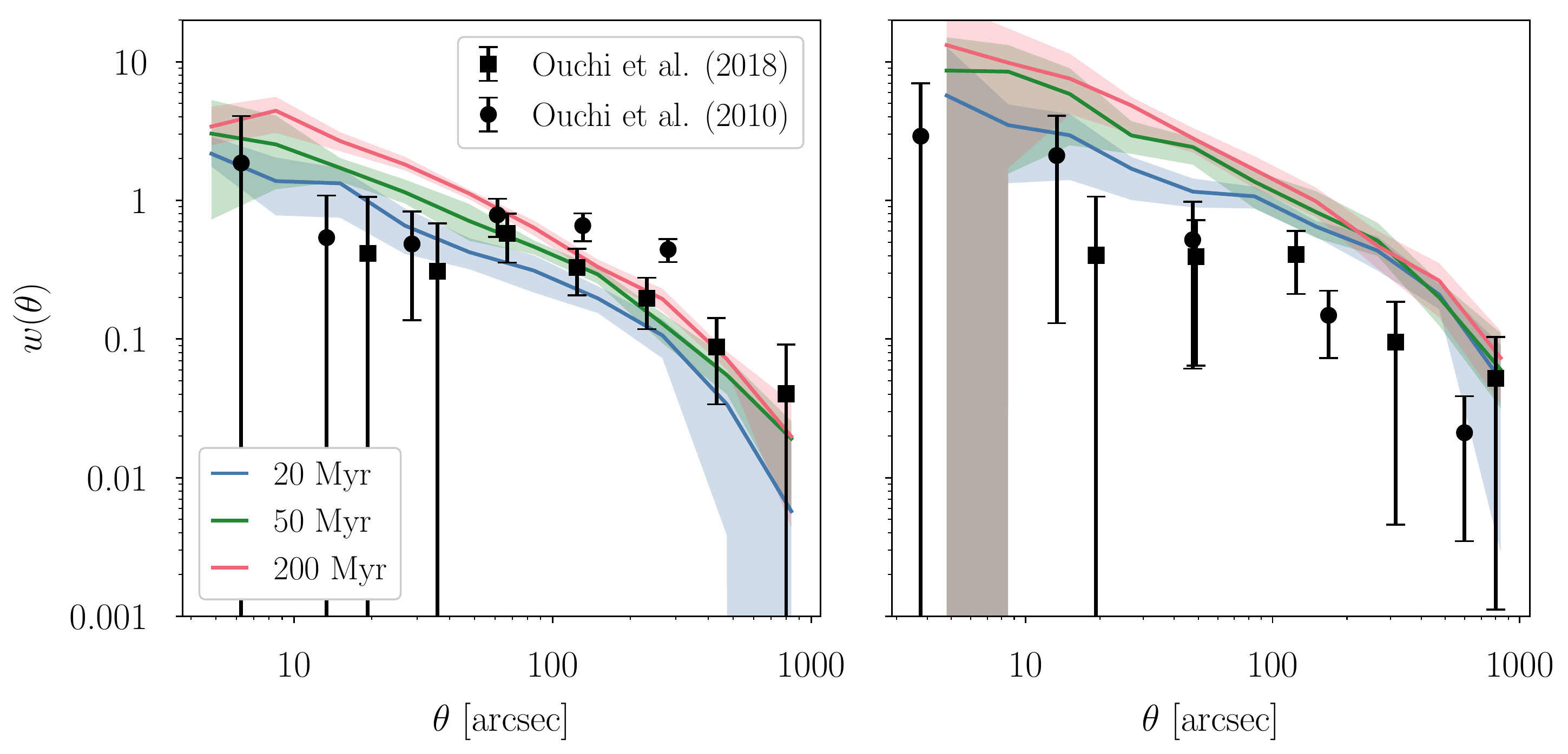}
    \caption{Angular correlation functions at $z=5.756$ and 6.604, comparing different
    choices of the $\Delta t$ duty cycle parameter: a smaller value of 20 Myr
    in \colb, our fiducial value of 50 Myr in \colg, and the fiducial value of
    \citet{2010ApJ...714L.202T} in \colr. All lines are calculated for the
    same reionization history.}
    \label{fig:vary_dt}
\end{figure*}

\subsection{Ly $\alpha$ luminosity function evolution}
\label{sec:LF}
Given the mock LAE sample for each redshift which includes the IGM transmission
fraction and the intrinsic luminosity, we can construct the observed sample
using Eq.~(\ref{eq:obsL}). This allows us to construct the (spatial) luminosity
function and compare to observed results. In Figure~\ref{fig:LF} we show the
evolution of the luminosity function for our six reionization histories,
using our fiducial $\Delta v \propto v_\mathrm{circ}$ model. 
From left to right the redshift of our mock population increases
from $z=5.756$ to $z=7.444$. As described above, the luminosity function is 
calculated by slicing the
simulation volume and taking the mean across the slices for each luminosity bin.
 The shading represents the 68\% scatter around this slice mean.

We see that the reionization history which qualitatively fits the observed 
data across the available narrowband redshifts is the Delayed Very Late model, 
suggesting that a later reionization is most consistent
\citep[as found in][]{2018arXiv180906374K,2018arXiv180706209P}. We note that the most
difficult constraint to match is the strong attenuation seen in the $z=7.3$ data.
The earlier reionization histories under-predict the IGM attenuation required 
to match the data at $z=7.3$. Since these datasets require the deepest
observations in order to find the very rare LAEs visible at such high
redshifts, it is also possible that some of the $z=7.3$ bins are not fully complete
and may move up in the future SILVERRUSH data release. 
Attenuation at $z=7.3$ maybe evidence for an even later start to reionization,
even later perhaps than our Very Late model.
We also note that there is some inconsistency between the different observed
datasets at $z=7.0$, which has been discussed in the literature.

\subsection{Clustering evolution}
As with the luminosity function calculation, we divide the simulation volume
into slices of approximately the same depth as the narrowband surveys (10 slices
of depth 32 cMpc/h), and assume the same luminosity cuts. We then use the 
\citet{1993ApJ...412...64L} estimator:
\begin{equation}
    w(\theta) = \frac{DD(\theta) - 2DR(\theta) + RR(\theta)}{RR(\theta)},
\end{equation}
to calculate the angular correlation function $w(\theta)$, where $DD(\theta)$ is
the number of galaxy-galaxy pairs at separation theta, 
$RD(\theta)$ the number of random-galaxy pairs, and $RR(\theta)$ the 
random-random pairs, all of which are normalized appropriately. 
We employ the \textsc{swot} code \citep{2012A&A...542A...5C} 
to perform the calculation efficiently. Our random field is generated 
by drawing from a uniform distribution, with similar number density to that
of \citet{2017arXiv170407455O} \citep[see][]{2018PASJ...70S...7C}. 
In Figure~\ref{fig:acf} we plot the angular correlation function for each
test redshift, showing the different reionization histories as in Figure~\ref{fig:LF}.
The scatter across the slices is shown by the shading,
whilst the lines are the mean value of $w(\theta)$. We note that at both $z=5.756$
(left) and $z=6.604$ (middle-left) our predictions are within the scatter of the 
observational results from \citet{2017arXiv170407455O}. In the $z=6.604$ panel,
we already start to see the effect of the different transmission fractions predicted
by the reionization history models. The Very Late model, with the lowest
average transmission, gives the highest clustering signal.

Considering the SILVERRUSH clustering data 
\citep{2017arXiv170407455O,2010ApJ...723..869O} alone, there is very little
evolution in the angular correlation function. Since we expect the clustering
to be increased due both to the higher bias at higher redshifts as well as due
to the ionized bubble structure of the IGM, this lack of evolution between 
$z=5.7$ and 6.6 is puzzling, perhaps suggesting that samples are not yet large 
enough for an accurate determination of the clustering of this higher redshift.
From the modelling perspective, it would be possible to reduce the 
predicted clustering signal at $z=6.6$ further by using a shorter duty cycle, 
i.e. a lower value of $\Delta t$, however this will also affect the quality of the
agreement at $z=5.7$.

\subsection{Clustering predictions for $z\geq7.0$}
In the right-hand two panels of Figure~\ref{fig:acf} we make predictions for the
clustering signal at redshifts $z=6.860$ and $=7.444$, which are close to the
narrowband filters NB973 and NB101. We see a similar pattern as was observed for
the left-hand panels: the clustering signal increases for all models, and in 
particular the HM12 history (in blue) exhibits the least clustering whilst the
Very Late model (in orange) exhibits the most. On intermediary scales these models
are non-overlapping at the 68\% scatter level. In particular the 2-point correlation
function is a function of the reionization history, suggesting that
measurements at these redshifts could be strongly constraining.

Apart from the angular correlation function, the clustering signal might also
be useful for understanding ionized bubble structure deep into reionization
\citep[e.g.][]{2016MNRAS.463.4019K}.
As an example of how this might be possible, we plot a projected mock LAE slice in 
Figure~\ref{fig:map} indicating the intrinsic population in grey and the observed
population in red, at $z=6.604$ for the Very Late reionization history. 
We also plot a contour partioning the map based on the projected ionization 
fraction, to indicate where the LAEs reside with respect to the ionized bubbles.
We note that the observed LAEs seem to lie within ionized bubbles, whereas the
unobserved objects are in neutral regions. This configuration could therefore
allow us to constrain bubble sizes, perhaps in concert with proposed 21 cm
methods \citep{2018MNRAS.473.2949G}. 
We leave the construction of such methods to future work.

\begin{figure}
    \includegraphics[width=\columnwidth]{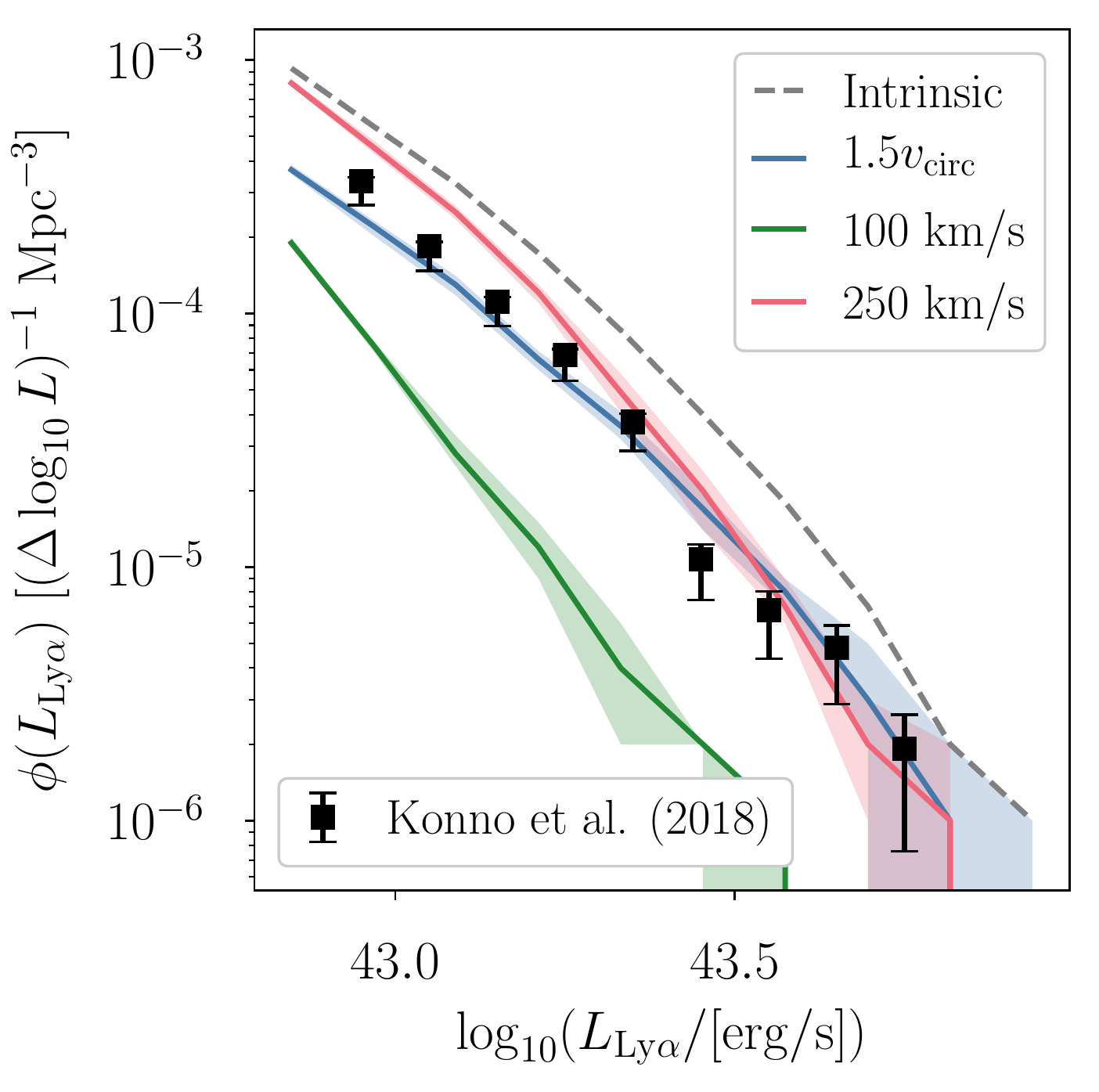}
    \caption{Luminosity functions at $z=5.756$, comparing different choices for
    the $\Delta v$ emission parameter: our fiducial 1.5 $v_\mathrm{vcirc}$
    in \colb, a fixed value of 100 km/s in \colg, and a fixed value of 250 km/s in
    \colr. All lines are calculated using the same reionization history.}
    \label{fig:vary_dv}
\end{figure}

\subsection{Differential evolution of the bright and faint end of the luminosity function}
In Figure~\ref{fig:brightfaint} we plot a 2D histogram for the mock LAE
population at $z=7.444$, binning by transmission fraction and `intrinsic'
Ly $\alpha$ luminosity. We see that for our chosen transmission model the
brighter LAEs are preferentially more visible than the fainter objects. In 
particular the mean and median transmission fractions are shown with the
red and white lines, and we see that both curve towards higher transmission
fractions as the luminosity increases.

Considering the results of \citet{2018MNRAS.tmp.1485W}, there are two components
of this differential visibility to understand. Firstly the absolute 
transmission fraction at a given redshift is strongly dependent on the 
emission profile; for our models it is therefore dependent on the choice of
$\Delta v$. Our choice in this work assumes $\Delta v \propto v_\mathrm{circ}$ which
means that $\Delta v \propto M_\mathrm{vir}^{\frac{1}{3}}$. This partly explains
the behaviour seen in Figure~\ref{fig:brightfaint}; other
theoretical work such as \citet{2018arXiv180101891M} assumed
$\Delta v \propto M_\mathrm{vir}$ and found a similar boost in the transmission
of bright LAEs. However in
\citet{2018MNRAS.tmp.1485W} it was found that there can also be a differential
visibility for the relative transmission fraction (i.e. the transmission fraction
at a given redshift relative to another redshift). The differential evolution
of this relative transmission was partly caused by the presence of brighter
LAEs within larger ionized regions, as well as the different host environments
of the haloes that the LAEs resided in. Importantly it was found that this
differential evolution of the relative transmission was insensitive to the 
choice of $\Delta v$.

\begin{figure}
    \includegraphics[width=\columnwidth]{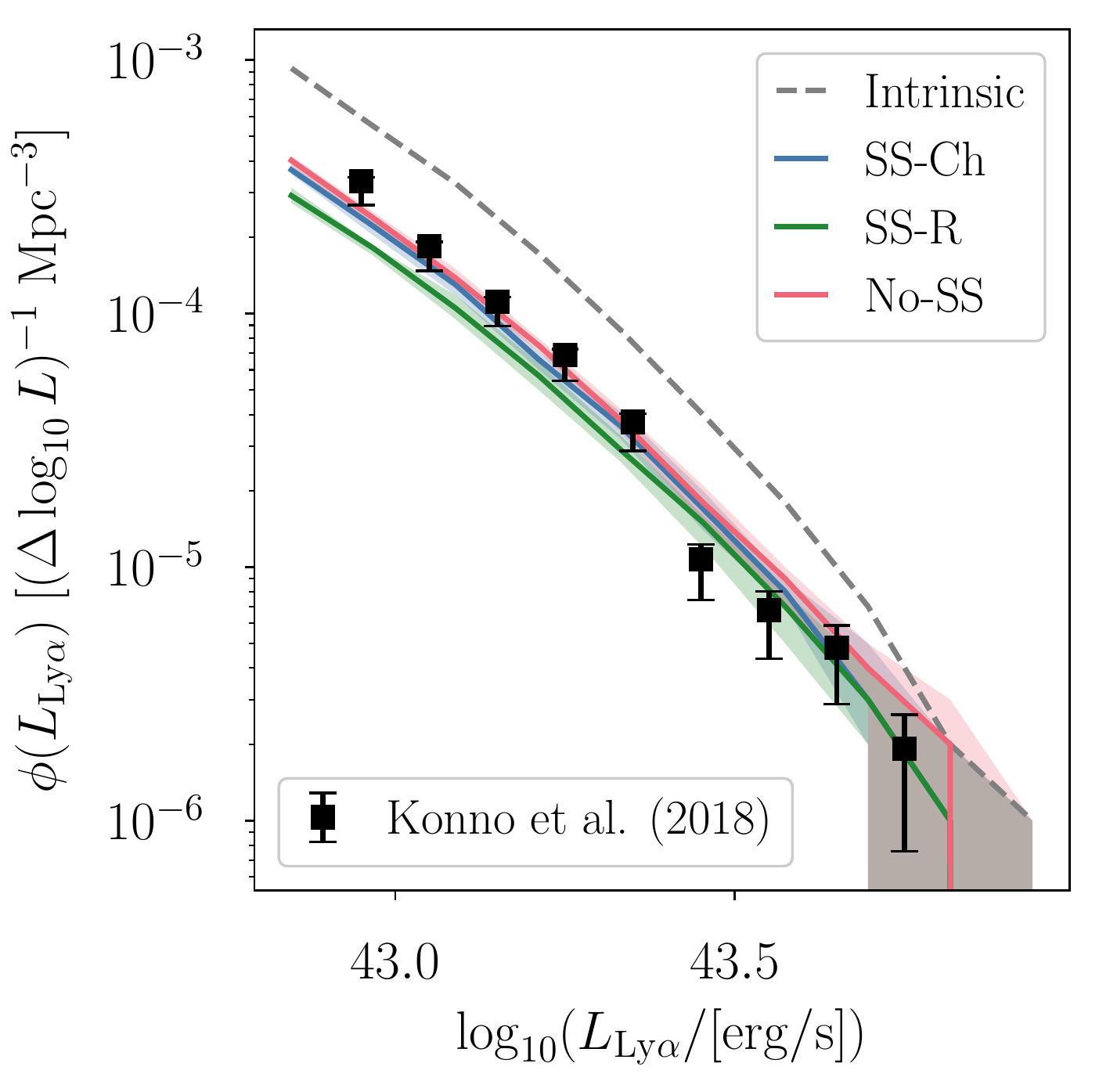}
    \caption{Luminosity functions at $z=5.756$, comparing different 
    self-shielding prescriptions: our fiducial \citet{2017arXiv170706993C} based choice
    in \colb, the prescription of \citet{2013MNRAS.430.2427R} in \colg, 
    and also using no self-shielding prescription in \colr.
    All lines are calculated using the same reionization history.}
    \label{fig:vary_ss}
\end{figure}

\subsection{Typical LAE masses}
In Figure~\ref{fig:massdist} we plot the distribution of host halo masses 
as a function of `observed' Ly $\alpha$ luminosity for our mock population at
$z=5.756$. Overplotted on our model histogram we show the observed NB816 data from 
\citet{2018arXiv181100556K} with a white marker. As in Figure~\ref{fig:brightfaint}
the mean and median of the distributions are shown in red and white lines.
We see the strong correlation between halo mass and luminosity, expected from
Eq~(\ref{eq:abun}), which prevails even after IGM attenuation.
The mean host halo mass of our $z=5.756$ mock sample is $10^{11.0} M_\odot$, whilst the
minimum is $10^{10.3} M_\odot$ and the maximum is $10^{12.6} M_\odot$.


\section{Discussion}
\label{sec:discussion}

Our model, detailed above, reproduces the evolution of the LAE luminosity
function and angular correlation function reasonably well. The main free 
parameters were chosen in our fiducial model as $\Delta v = 1.5 v_\mathrm{circ}$
and $\Delta t = 50$ Myr. 
We now discuss further the motivation for these choices,
and the effect of varying these parameters on the observables.

\subsection{The effect of varying $\Delta t$ on the clustering}
\label{sec:dt}
In Figure~\ref{fig:vary_dt} we show the the angular correlation function at
$z=5.756$ and 6.604 for three different values of $\Delta t$: 20, 50 and 200 Myr. We see
that increasing this parameter causes an increase in the clustering power, 
especially at smaller scales. In particular comparing to observations from
\citet{2017arXiv170407455O,2010ApJ...723..869O}, 
the 200 Myr duty cycle causes too much small 
scale correlation to be consistent with the observed correlation at scales 
around $10 \lesssim \theta \lesssim 60$ arcsec.

As discussed in section~\ref{sec:LBGs}, the $\Delta t$ parameter controls the
LBG duty cycle, weighting the abundance matching step towards haloes that have
undergone a change in mass within the past $\Delta t$ epoch. The motivation for
this prescription is that such variation might correlate with recent bursts of
star formation, and therefore UV luminosity and observability. 
\citet{2010ApJ...714L.202T} chose as their fiducial value $\Delta t = 200$ Myr, however
we see that in our implementation this does not match the SILVERRUSH 
clustering signal. We also note that LAE selected galaxies tend to have younger ages
than LBG selected populations \citep{2007ApJ...671..278G}.

We note again that the absence of any evolution in the observational data across
these redshifts is puzzling. Even the $\Delta t = 20$ Myr duty cycle model does
not achieve a low enough clustering to match the observations at $z=6.6$.

We find that the luminosity function is insensitive to these tested variations 
in $\Delta t$.

\subsection{The effect of varying $\Delta v$ on the luminosity function}
\label{sec:dv}
In Figure~\ref{fig:vary_dv} we show the luminosity function at $z=5.756$, and
the effect of varying the $\Delta v$ parameter. As the transmission fraction
is a strong function of the intrinsic emission profile 
\citep{2011MNRAS.414.2139D}, 
the luminosity function is also dependent on 
our assumptions about this profile. We see that a fixed value of $\Delta v$
does not fit well; for example the use of $\Delta v = 250$ km/s shown in 
\colr~does fit the bright end reasonable well but overpredicts the number density of
LAEs at the faint end. Fixing to a lower value, for example the fiducial choice
of \citet{2018MNRAS.tmp.1485W} of $\Delta v = 100$ km/s, results in too much 
attenuation by the IGM at all luminosities. Our fiducial choice in this work was
to set $\Delta v \propto v_\mathrm{circ}$, which means that it scales with the
LAE host halo mass as $\propto M_h^{1/3}$. Due to our population modelling
(described in section~\ref{sec:LBGs}--\ref{sec:transmission}) 
the LAE luminosity should scale with the
UV luminosity (with some stochasticity due to the REW distribution), and hence
the host halo mass. This leads to the transmission distribution seen in
Figure~\ref{fig:brightfaint} and the good agreement in Figure~\ref{fig:vary_dv}.

\subsection{The effect of self-shielding on the luminosity function}
\label{sec:SS}
Finally in  Figure~\ref{fig:vary_ss} we show the luminosity function at 
$z=5.756$, and the effect of varying the self-shielding prescription. Here we
compare our fiducial self-shielding prescription with the cases of no
self-shielding or a \citet{2013MNRAS.430.2427R} prescription. 
We see that our luminosity function
predictions are largely insensitive to the choice of prescription, although 
there is some increase in attenuation when self-shielding is added. We note
that the non-negligible attenuation is present even in the absence of 
self-shielding; this is due to the non-zero neutral fraction that is found in
photoionization equilibrium at the outskirts of the halo.


\section{Conclusions}
\label{sec:conclusions}
In conclusion we have built an empirically constrained, self-consistent model
of the evolution of LAEs in the epoch of reionization. 
This modelling made use of the halo population in the state-of-the-art
Sherwood simulations, and the hydrodynamic gas structure for quantifying the
transmission of Ly $\alpha$ emission through the IGM.
\begin{itemize}
\item We used the best fit REW probability distribution of 
\citet{2012MNRAS.419.3181D} to model the intrinsic REW distribution of 
our mock LAE population, and found that incorporating the IGM transmission 
fraction (calculated for each LAE individually) we reproduced the observed 
$z=5.7$ REW distribution. Our
transmission modelling assumes that the intrinsic LAE
emission is a single Gaussian peak, with a velocity offset proportional to
the host halo virial circular velocity. This gives a transmission fraction
probability distribution across the mock population which corrects
the overabundance of high equivalent widths 
predicted by the \citet{2012MNRAS.419.3181D} distribution at this redshift.

\item In both our original reionization histories and the delayed-end versions,
there is sufficient neutral hydrogen in the CGM and infalling gas further surrounding the
halo at $z\sim5.7$ that there is some attenuation redwards of Ly $\alpha$. The
attenuation by the neutral hydrogen in the CGM dominates
(near the end of reionization) over the attenuation due to the large-scale IGM, 
even with the presence of residual neutral islands in the delayed-end models 
(when the average neutral fraction is non-zero). However at higher redshifts
around the midpoint of reionization (when $Q \lesssim 0.5$) we find that
variations in the neutral fraction of the wider IGM have a dominant effect 
on the Ly $\alpha$ transmission compared to the CGM neutral gas.

\item Using this model we generated mock LAE populations at the redshifts of
interest for narrowband surveys, and made predictions for the luminosity 
function and angular correlation function. Comparing these predictions with
current data, in particular from the SILVERRUSH survey, we find that a rather late
reionization history (our Very Late model) is in best agreement.

\item In order to match the luminosity function across redshifts, we find that
the delayed-end Very Late model has the best fitting evolution whilst still
able to attenuate the signal enough at the highest redshifts.

\item Employing a duty cycle in our LBG modelling allowed us to match the
LAE 2-point correlation function with our mock population at $z=5.7$. Our predictions
for higher redshifts suggest that the ionization structure of the IGM can enhance
the clustering signal significantly already at $z=6.6$. The lack of evolution in the
current observed clustering at these redshifts is difficult to explain consistently
with the evolution in the luminosity function and equivalent width distribution,
and may suggest that the clustering at this redshift has not been measured with
sufficient accuracy to extract the effect of reionization.

\item In agreement with the results of \citet{2018MNRAS.tmp.1485W}, which found 
that comparing more and less massive host haloes there is a differential evolution in the 
relative transmission fraction (e.g. $T_\mathrm{IGM}(z)/T_\mathrm{IGM}(z=5.7)$),
we find that our transmission model also leads to a difference in the absolute
transmission (e.g. $T_\mathrm{IGM}(z)$). 
We find that the more luminous LAEs are preferentially less
attenuated by the IGM neutral fraction, albeit with a large scatter.
\end{itemize}

Lyman-$\alpha$ emitting galaxies have been considered as probes of reionization
for over 20 years, and many attempts have been made
at observing and modelling their behaviour at high redshifts. The ongoing
ambitious Ly $\alpha$ surveys are starting to collect samples sufficiently large
to allow us to put tight contraints on the reionization history of hydrogen. 
We find that the evolution
of the luminosity function and angular correlation function are indeed 
strongly dependent on the reionization history, such that further observations
at $z\geq7$ and future Ly $\alpha$ surveys should allow us to map out in detail
the second half ($Q \geq 0.5$) of reionization.


\section*{Acknowledgements}
We thank the referee Akio Inoue for his detailed comments and suggestions
for improving the manuscript.
We thank Masami Ouchi, Takatoshi Shibuya and Ryohei Itoh for providing their
data-points to allow us to compare our predictions with their observations.
We also thank Masami for his comments on the initial manuscript.
We acknowledge useful discussion with Prakash Gaikwad,  
Ewald Puchwein, Koki Kakiichi, Laura Keating and Tirth Choudhury.
LHW is supported by the Science and Technology Facilities Council
(STFC). 
Support by ERC Advanced Grant 320596 `The Emergence of Structure 
During the Epoch of Reionization' is gratefully acknowledged.  We acknowledge
PRACE for  awarding  us  access  to  the  Curie  supercomputer,  based in
France  at  the  Tr\'{e}s  Grand  Centre  de  Calcul  (TGCC). 
This work was performed using the Cambridge Service for Data Driven Discovery 
(CSD3), part of which is operated by the University of Cambridge Research 
Computing on behalf of the STFC DiRAC HPC Facility (www.dirac.ac.uk). 
The DiRAC component of CSD3 was funded by BEIS capital funding via STFC 
capital grants ST/P002307/1 and ST/R002452/1 and STFC operations 
grant ST/R00689X/1. DiRAC is part of the National e-Infrastructure.
This work made use of the \texttt{SciPy} \citep{SCIPY} ecosystem of libraries
for Python including: \texttt{NumPy} \citep{NUMPY}, \texttt{Matplotlib}
\citep{MATPLOTLIB} and \texttt{Cython} \citep{CYTHON}.



\bibliographystyle{mnras}
\bibliography{references}



\appendix
\section{Comparing the use of full radiative transfer post-processing with the
excursion set based method}
\label{appendix:compareRT}

\begin{figure*}
    \includegraphics[width=2\columnwidth]{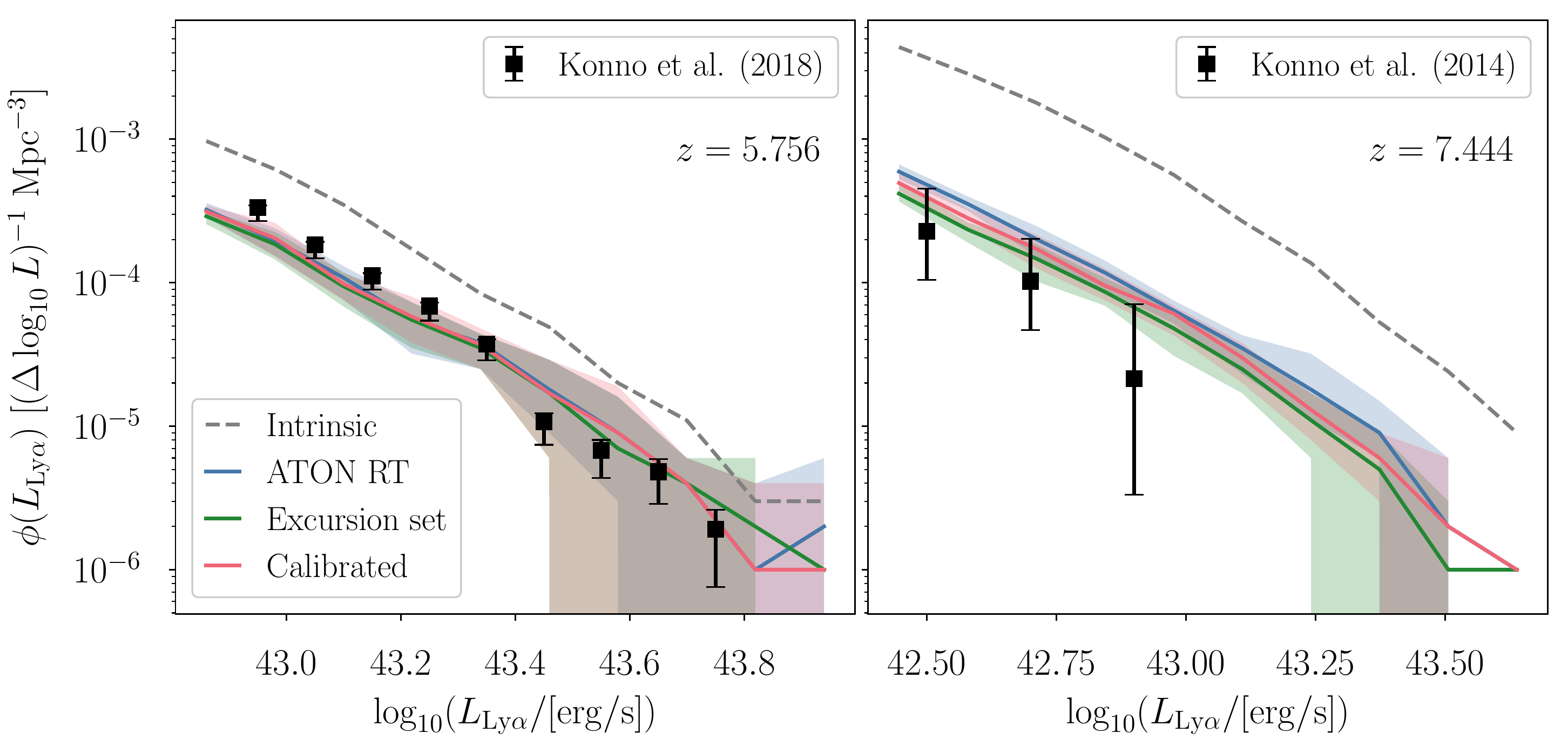}
    \caption{Luminosity function at $z=5.756$ and 7.444 for the same 
    reionization history: calculated using full radiative transfer 
    post-processing in \colb, a
    simple excursion set implementation in \colg, and
    using our calibrated excursion set method in \colr.}
    \label{fig:compareRTl4l}
\end{figure*}

In this section we compare three post-processing methods for constructing the 
large-scale ionization field within the hydrodynamic simulation. We compare 
(i) a simple excursion set prescription, (ii) the calibrated excursion set method used 
in \citet{2015MNRAS.452..261C} (detailed in
\citet{2018MNRAS.tmp.1485W}) and also in this work, 
and (iii) a full radiative transfer calculation. The full radiative transfer post-processing
was performed using the \textsc{Aton} code \citep{2008MNRAS.387..295A}, 
as detailed in \citet{2018arXiv180906374K}. All methods were applied to the
grids of the $L=160$ cMpc/h, $N = 2 \times 2048^3$ Sherwood simulations 
to generate the ionized fraction in each grid cell. So that we are
comparing like-for-like, we apply the same self-shielding prescription
for all three methods to model the small-scale ionization structure.
We then generated mock
LAE populations as detailed in sections~\ref{sec:LBGs}--\ref{sec:transmission} 
and calculated
the observable luminosity function and equivalent width distribution.

In Figure~\ref{fig:compareRTl4l} we compare the luminosity functions
predicted by these three methods.
We show the \textsc{Aton} reionization model of 
\citet{2018arXiv180906374K} in \colb, the equivalent reionization history
implemented with the simple excursion set method in \colg, and finally
our calibrated excursion set method in \colr. 
In order to compare like-for-like we
have applied the excursion set method in the following two ways: firstly
in the simple prescription we have used as inputs the mass-averaged ionized 
fraction, $\langle x_\mathrm{HII} \rangle_\mathrm{m}$, and the 
volume-averaged photoionization rate within ionized regions, 
$\langle \Gamma_\mathrm{HI} \rangle_\mathrm{v}$, from the \textsc{Aton} fields.
Given these two inputs we can apply the excursion set method and the 
self-shielding prescription to create the ionization structure at large and
small scales. Secondly we have used our calibrated method, in which we take
the  mass-averaged ionized 
fraction as before, but instead of using the \textsc{Aton} photoionization rate 
directly we rather take the mean free path, $\lambda_\mathrm{mfp}$, 
from the \textsc{Aton} fields.
This is used to solve for the background photoionization rate consistently within
our simulation volume (as detailed in \citet{2018MNRAS.tmp.1485W}, section 2.3).

We find that the three methods give similar predictions, consistent within the 68\%
scatter across the slices, and that the calibrated method is closer to the
full \textsc{Aton} method at all redshifts. When the photoionization rate
is $\Gamma_\mathrm{HI}\gtrsim 10^{-13} \mathrm{s}^{-1}$ the simple excursion set
model is close to the other models, however they start to diverge
at higher redshifts when this is no longer the case. We find that the simple excursion
set method, which assumes a uniform UV background, slightly overattenuates the luminosity
function compared to the \textsc{Aton} method. We note that the full radiative
transfer will not have a uniform UV background, and instead we would expect 
higher photoionization rates near to the LAEs (where also the gas density is highest).
This means that when we compare the neutral hydrogen densities around the LAEs,
the \textsc{Aton} method gives a more ionized CGM compared to the excursion set
method which sees the uniform UV background. This can be seen for $z=7.444$ in
Figure~\ref{fig:compareRTl4l}. For the lower redshifts near to the end of
reionization ($z\sim6$) we find that the photoionization rate is high enough
that the ionized fraction saturates, such that the two methods are in good
agreement, as seen in the left hand panel. 
In general our calibrated method predicts slightly higher background 
photoionization rates which improves the agreement with the \textsc{Aton} 
results compared to the simple case.

\section{Luminosity dependence of the CGM and IGM attenuation}
\label{appendix:compareCGM}

\begin{figure*}
    \includegraphics[width=2\columnwidth]{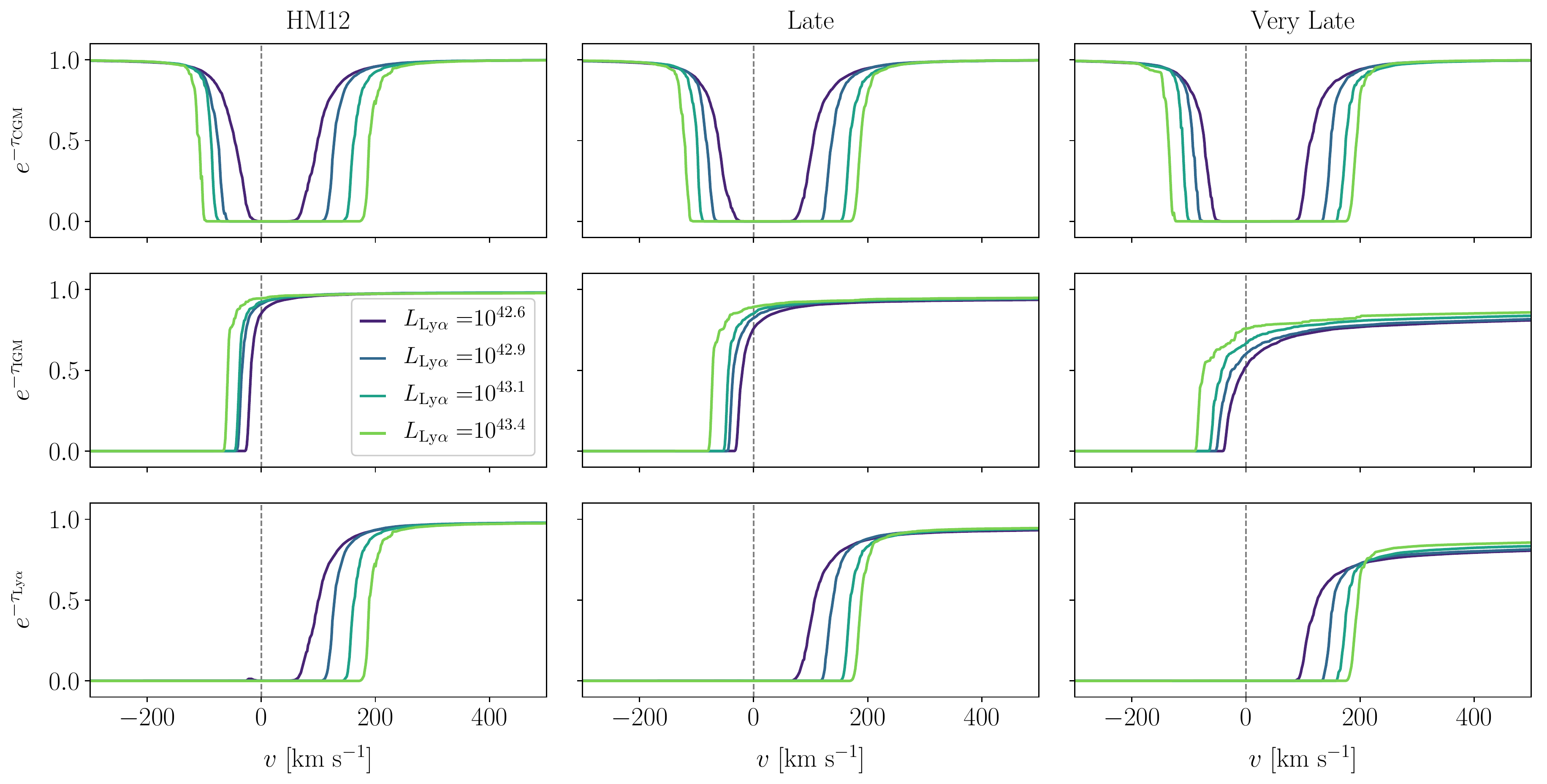}
    \caption{The median transmission in different luminosity bins for the
    mock LAE population at $z=7.444$, split into CGM (top), IGM (middle) and
    total (bottom) components as in Figure \ref{fig:components}. 
    The different (logarithmic) luminosity bins are indicated
    by the colour of the lines, with the centre of the bin indicated in the legend.
    From left to right we show the three original reionization histories: HM12,
    Late and Very Late.}
    \label{fig:components_alt}
\end{figure*}

In section~\ref{sec:CGM} we demonstrated that the CGM can play a significant role
in the attenuation of Ly $\alpha$, particularly relevant near the end of
reionization. However in that section we considered only the median transmission
of the full observed sample. We now quantify the variation in the role of CGM/IGM 
components across luminosity. We note that the Ly $\alpha$ luminosity of our 
sample broadly scales with the host halo mass (see Figure~\ref{fig:massdist}),
hence the following also applies to the variation with mass. These results
confirm what was found in \citet{2018MNRAS.tmp.1485W}.

In Figure~\ref{fig:components_alt} we show the median transmission curves,
split into CGM, IGM and total components as in Figure~\ref{fig:components}, but 
calculated for samples in different luminosity bins. For brevity we show only
the $z=7.444$ original reionization history curves, 
from left to right showing the HM12, Late and Very Late
models respectively. The colour scale of the lines indicates the different 
luminosity bins, with the faintest bin centred on $10^{42.6}$ erg/s and the
brightest centred on $10^{43.4}$ erg/s.

We see in Figure~\ref{fig:components_alt} that the attenuation from both the CGM and IGM 
components depends on luminosity. In particular for the CGM component
we see that the fainter objects have more transmission around line-centre,
compared to the brightest objects which have a wider absorption trough.
This results from higher densities around the more massive haloes, which for the
same UV background leads to more neutral gas around the brighter LAEs compared
to the faint ones. In contrast for the IGM component, the brighter LAEs have
more transmission compared to the faint ones. This is because the brighter LAEs
(more massive haloes) reside in larger ionized regions compared to the fainter
LAEs. As found in \citet{2016MNRAS.463.4019K}, these different luminosity 
dependences work against each other in the combined transmission. 

Considering a given luminosity bin (one of the colours in Figure~\ref{fig:components_alt})
we see that the CGM component is similar across the different reionization history
models. In comparison the IGM component shows more difference, with the Very Late
model in particular showing the most attenuation. This suggests that even with
a CGM component that depends on the background photoionization rate, for low
enough average ionized fractions (at high enough redshifts into the second
half of reionization) the LAE transmission is indeed a strong function of the 
IGM neutral fraction, and hence LAE observations can provide good constraints.

We note finally that our modelling of the CGM makes various simplifying assumptions
which may start to break down in the brightest LAEs. In particular
we do not model the source emissivity when calculating the neutral hydrogen
density around the halo. Furthermore with our assumption of an intrinsic emission 
profile we ignore the gas within $R_\mathrm{vir}$ of the halo centre. These 
limitations may affect the CGM transmission for the most massive (brightest)
haloes.

\section{Reionization history parameters}
\label{appendix:Qtable}

\begin{figure}
    \includegraphics[width=\columnwidth]{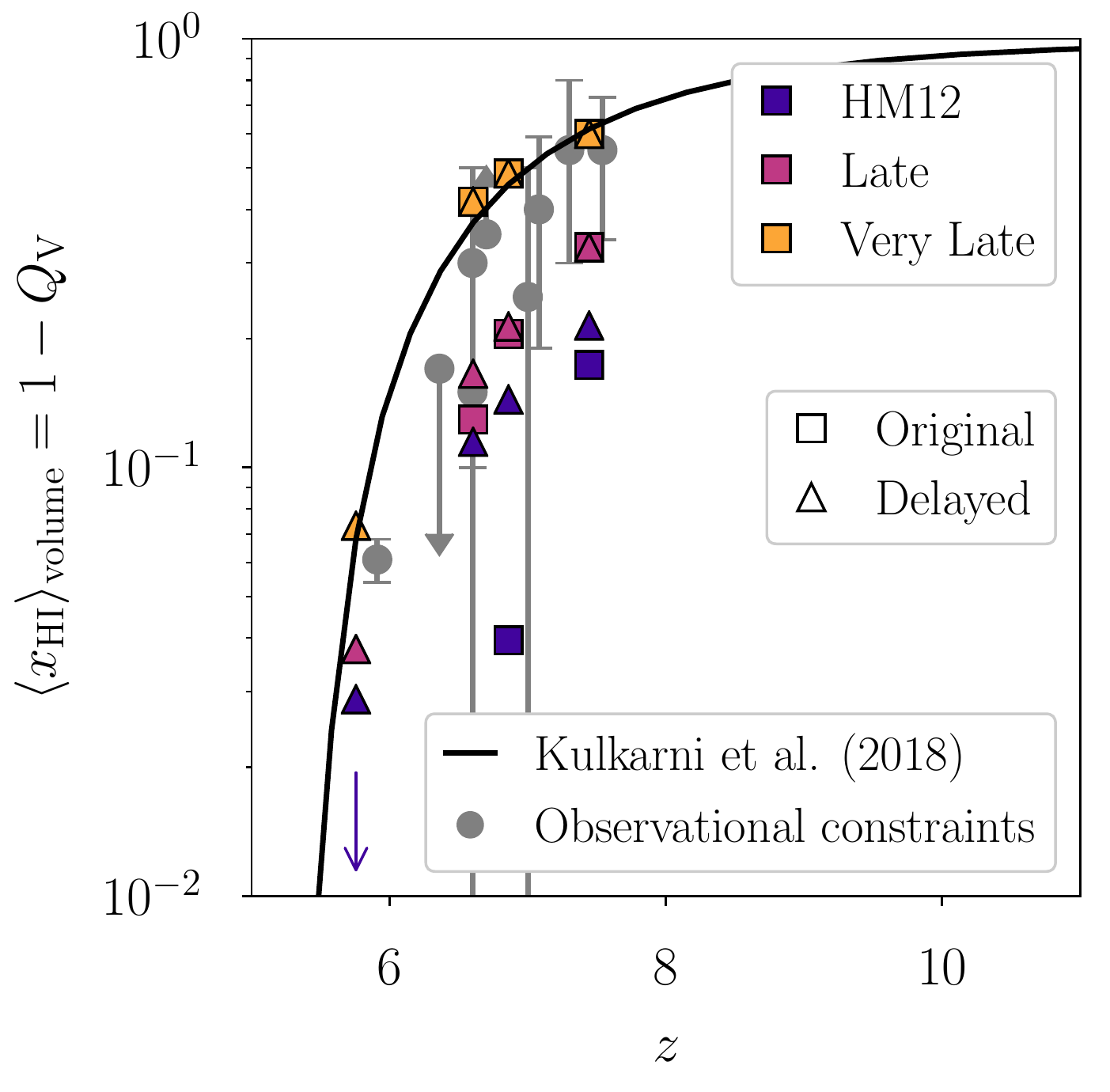}
    \caption{The volume-averaged neutral fraction predicted in our models shown
	in squares (original histories) and triangles (delayed-end histories), compared to 
    recent observational constraints from a variety of studies, shown with grey circles.
    These correspond to measurements of LAEs \citep[][$z=6.6$]{2017arXiv170501222K},
    \citep[][$z=7.3$]{2014ApJ...797...16K}, \citep[][$z=6.6$]{2017arXiv170407455O}, 
	\citep[][$z=7$]{2018arXiv180505944I}; QSOs \citep[][$z=7.1$]{2017MNRAS.466.4239G},
	\citep[][$z=7.5$]{2018Natur.553..473B}; and GRBs \citep[][$z=5.9$]{2016PASJ...68...15T},
	\citep[][$z=6.4$]{2006PASJ...58..485T}, \citep[][$z=6.7$]{2009ApJ...693.1610G}.
	The black solid line shows the model of \citet{2018arXiv180906374K}, which is
	similar to our delayed-end Very Late model, and was found to reproduce the
    opacity fluctuations in the Ly $\alpha$ forest. 
	The vertical scale of this figure is logarithmic.}
    \label{fig:obsQ}
\end{figure}

\begin{table*}
\caption{Volume-averaged neutral fractions,
$\langle x_\mathrm{HI} \rangle_\mathrm{v}$,
for the reionization histories considered in this work.}
\begin{tabular}{c|c|c|c|c|c|c}
 \hline
     &      & Original &   &   &  Delayed & \\ \hline
 $z$ & HM12 & Late & Very Late & HM12 & Late & Very Late \\ \hline 
 5.756 & 0.0000 & 0.0000 & 0.0000 & 0.0288 & 0.0377 & 0.0731\\ \hline
 6.604 & 0.0000 & 0.1294 & 0.4168 & 0.1148 & 0.1654 & 0.4168\\ \hline 
 6.860 & 0.0395 & 0.2049 & 0.4849 & 0.1440 & 0.2132 & 0.4849\\ \hline
 7.444 & 0.1734 & 0.3263 & 0.5999 & 0.2142 & 0.3263 & 0.5999\\ \hline
\end{tabular}
\label{tab:Qpredictions}
\end{table*}

\begin{table*}
\centering
\caption{Calibrated background photoionization rates,
$\log_{10}(\Gamma_\mathrm{HI}/\mathrm{s}^{-1})$,
for the reionization histories considered in this work.}
\begin{tabular}{c|c|c|c|c|c|c}
 \hline
     &      & Original &   &   &  Delayed & \\ \hline
 $z$ & HM12 & Late & Very Late & HM12 & Late & Very Late \\ \hline 
 5.756 & -12.50 & -12.61 & -12.64 & -13.45 & -13.47 & -13.32\\ \hline
 6.604 & -12.79 & -13.04 & -13.26 & -13.26 & -13.24 & -13.26\\ \hline 
 6.860 & -12.88 & -13.14 & -13.26 & -13.25 & -13.21 & -13.26\\ \hline
 7.444 & -13.07 & -13.21 & -13.27 & -13.30 & -13.20 & -13.27\\ \hline
\end{tabular}
\label{tab:Gpredictions}
\end{table*}

Our reionization prescription takes as input a reionization history given in
terms of the mass-averaged ionized fraction evolution with redshift. 
However observers usually infer the volume-averaged fraction, which is
weighted more towards volume-filling voids. In order to convert between the 
two quantities in practice is difficult; in the case of our simulations we
can use a given snapshot (which is a realization of the mass distribution) in order
to measure both quantities.

In Table~\ref{tab:Qpredictions} we show for each of our reionization history 
models the values of the volume-averaged neutral fraction, and similarly in
Table~\ref{tab:Gpredictions} the photoionization rates, at
the redshifts we considered in this work. In Figure~\ref{fig:obsQ} we compare 
these values to a selection of observational constraints on the 
average neutral fraction of the IGM, on a logarithmic scale.
These constraints were derived from observations of 
LAEs \citep{2017arXiv170501222K,2014ApJ...797...16K,2017arXiv170407455O,
2018arXiv180505944I}, QSOs
\citep{2017MNRAS.466.4239G,2018Natur.553..473B} 
and GRBs \citep{2016PASJ...68...15T,2006PASJ...58..485T,2009ApJ...693.1610G}. 
We note that the points representing our reionization histories
are not constraints but predictions of the models. The squares show the
evolution in our original reionization histories, whilst the triangles show
the delayed-end histories. Note that the delayed-end histories only deviate
near the end of reionization, so for example the delayed-end Very 
Late model has the same neutral fraction evolution as the original for $z>6$.
The arrow indicates that because the original models have reionized by $z=6$ 
(Late/Very Late) or $z=6.7$ (HM12), the
lower redshift datapoints where $\langle x_\mathrm{HI} \rangle_\mathrm{v} = 0$
are not visible within the logarithmic scale of the figure.
We also show with a solid black curve the reionization history model of 
\citet{2018arXiv180906374K} which was able to match 
opacity fluctuations in the Ly $\alpha$ forest, and note that it
is very similar to our successful delayed-end Very Late model. 
In particular we highlight that both versions of our Very Late model are
consistent with the SILVERRUSH observational constraints.


\bsp	
\label{lastpage}
\end{document}